\documentclass[apj,iop,twocolumn,numberedappendix,appendixfloats,twocolappendix]{emulateapj}
\usepackage{natbib}
\usepackage{graphicx}
\usepackage{bm}
\usepackage{amssymb}
\usepackage[colorlinks,citecolor=blue,linkcolor=blue,urlcolor=blue]{hyperref}

\newcommand{\be}{\begin{equation}}
\newcommand{\ee}{\end{equation}}
\newcommand{\bea}{\begin{eqnarray}}
\newcommand{\eea}{\end{eqnarray}}

\begin{document}

\title{Distortion of a relativistic jet echoing a magnetic flux eruption}
\shorttitle{Post-eruption jet distortion}

\author{
\vskip -4em
Krzysztof~Nalewajko$^{\rm 1}$,
Mateusz~Kapusta$^{\rm 2}$,
Bart~Ripperda$^{\rm 3,4,5,6}$
\&~Alexander~A.~Philippov$^{\rm 7}$}
\shortauthors{Nalewajko, Kapusta, Ripperda \& Philippov}

\affiliation{
$^{\rm 1}$
Nicolaus Copernicus Astronomical Center, Polish Academy of Sciences, Bartycka 18, 00-716 Warsaw, Poland
\\
{\tt knalew@camk.edu.pl}
\\
$^{\rm 2}$
Astronomical Observatory, University of Warsaw, Al. Ujazdowskie 4, 00-478 Warsaw, Poland
\\
$^{\rm 3}$
Canadian Institute for Theoretical Astrophysics, 60 St. George St, Toronto, Ontario M5S 3H8, Canada
\\
$^{\rm 4}$
Department of Physics, University of Toronto, 60 St. George St, Toronto, ON M5S 1A7, Canada
\\
$^{\rm 5}$
David A. Dunlap Department of Astronomy, University of Toronto, 50 St. George St, Toronto, ON M5S 3H4, Canada
\\
$^{\rm 6}$
Perimeter Institute for Theoretical Physics, 31 Caroline St N, Waterloo, ON, N2L 2Y5, Canada
\\
$^{\rm 7}$
Department of Physics, University of Maryland, College Park, MD 20742, USA
}

\begin{abstract}
Magnetized accretion onto spinning black holes can accumulate a large magnetic flux across the event horizon and launch a pair of relativistic jets via the Blandford-Znajek mechanism.
In the magnetically saturated (arrested) state, excess magnetic flux is ejected from the black hole in episodic magnetic flux eruptions, which result in a significant yet temporary reduction of jet power.
We analyze results of a high-resolution 3D general-relativistic magneto-hydro-dynamic numerical simulation of geometrically thick magnetically saturated accretion onto a high-spin Kerr black hole for a single cycle of magnetic flux eruption and accumulation.
We show that following an eruption, a weakened jet develops a strong helical distortion with distinct structure of magnetic fields -- the poloidal field along the jet core is unaffected by the eruption;
while toroidal field lines, ejected from the black hole during the eruption and later re-advected onto it, form poloidal `bypasses' along the inner jet sheath.
Such a distortion may appear in sources fed by geometrically thick accretion flows as an asymmetric superluminal knot, strongly interacting with the jet sheath along an oblique working surface.
The jet section re-powered by magnetic flux re-accumulated on the black hole is tilted by a few degrees, implying significant variations in radiation boost towards observers of BL Lac blazars.
The intrinsic structure of the jet spine is consistent with axisymmetric semi-analytical models.
\end{abstract}

\keywords{
accretion
--- black hole physics
--- magnetic fields
--- magnetohydrodynamics (MHD)
--- relativistic jets
}

\section{Introduction}

Relativistic jets are spectacular manifestations of many accreting black holes (BHs).
These powerful collimated outflows are observed in active galactic nuclei (AGN, including blazars and radio galaxies; \citealt{2016ARA&A..54..725M,2019ARA&A..57..467B}), stellar X-ray binaries (XRBs, including microquasars; \citealt{1999ARA&A..37..409M}), and gamma-ray bursts (GRBs; \citealt{2004RvMP...76.1143P}).
Modeling of the observed very high broad-band luminosities \citep[e.g.,][]{2014ApJ...789..161N,2016ApJ...824L..20A,2023SciA....9I1405O} requires for the jets to sustain very efficient energy conversions: rotational to magnetic ($\mathcal{E}_{\rm rot} \to \mathcal{E}_{\rm mag}$; to power them), magnetic to (bulk-)kinetic ($\mathcal{E}_{\rm mag} \to \mathcal{E}_{\rm kin}$; to accelerate them), magnetic and/or kinetic to internal (random-kinetic) ($\mathcal{E}_{\rm mag},\mathcal{E}_{\rm kin} \to \mathcal{E}_{\rm int}$; to accelerate particles), internal to radiative ($\mathcal{E}_{\rm int} \to \mathcal{E}_{\rm rad}$; to explain observations).
In principle, the $\mathcal{E}_{\rm rot} \to \mathcal{E}_{\rm mag}$ conversion can be explained by the centrifugal jet launching mechanisms \citep[e.g.,][]{1976MNRAS.176..465B},
the $\mathcal{E}_{\rm mag} \to \mathcal{E}_{\rm kin}$ conversion by the magnetic nozzle mechanism \citep[e.g.,][]{1989ASSL..156..129C},
the $\mathcal{E}_{\rm int} \to \mathcal{E}_{\rm rad}$ conversion by standard non-thermal radiative mechanisms \citep{1986rpa..book.....R}.
The $\mathcal{E}_{\rm mag},\mathcal{E}_{\rm kin} \to \mathcal{E}_{\rm int}$ conversion, the problem of efficient dissipation, remains elusive \citep{2015MNRAS.450..183S}.

Stationary axisymmetric models of relativistic jets, supported by general-relativistic numerical simulations, can account for conversions $\mathcal{E}_{\rm rot} \to \mathcal{E}_{\rm mag}$ and $\mathcal{E}_{\rm mag} \to \mathcal{E}_{\rm kin}$;
conversion $\mathcal{E}_{\rm kin} \to \mathcal{E}_{\rm int}$ can in principle be realized by reconfinement shocks with modest efficiencies even in the hydrodynamical limit \citep{2009MNRAS.392.1205N}.
Departures from stationarity or axisymmetry received less theoretical attention despite clear observational evidence in the form of, e.g.,
time variability of radiation at all wavelengths \citep[e.g.,][]{2009ApJ...699..305H,2022ApJ...927..214G},
or direct imaging of asymmetric structures across jets \citep[e.g.,][]{2021ApJ...923L...5P,2023NatAs...7.1359F}.
A non-stationary jet engine has been considered for GRB \citep[e.g.,][]{1994ApJ...430L..93R} and blazars \citep[e.g.,][]{2001MNRAS.325.1559S} in the hydrodynamical limit, leading to remote dissipation via internal shocks.
In the inner jet regions dominated by magnetic energy density, instead of shocks one can expect magnetic dissipation involving reconnection of magnetic field lines \citep[e.g.,][]{2009MNRAS.395L..29G,2011MNRAS.413..333N,2011ApJ...726...90Z,2012MNRAS.419..573M}, which may also be required for efficient $\mathcal{E}_{\rm mag} \to \mathcal{E}_{\rm kin},\mathcal{E}_{\rm int}$ conversion \citep{2019MNRAS.484.1378G}.

In recent years, it has been recognized that the power of relativistic jets launched from spinning and accreting BHs by the Blandford-Znajek mechanism \citep{1977MNRAS.179..433B} can be strongly modulated due to a saturation mechanism for magnetic flux collected across the BH horizon scaled by the mass accretion rate, $\Phi_{\rm BH} / \dot{M}_{\rm acc}^{1/2}$.
When this parameter exceeds a critical value of $\sim 50$ \citep{2011MNRAS.418L..79T}, the pressure of accretion flow is unable to keep the magnetic flux on the BH horizon, allowing for large-scale reconnection between two magnetic hemispheres driving powerful eruptions \citep{2020ApJ...900..100R,2022ApJ...924L..32R}.
Such accretion has been described to be in the MAD (magnetically arrested disk) state \citep{2003PASJ...55L..69N},
although a true MAD with circularized magnetic barrier can be realized in geometrically thin disks \citep{2022ApJ...935L...1L,2022cosp...44.1769M};
for recent discussions see \cite{2024A&A...692A..37N,2025arXiv251025842J}.
A large eruption can temporarily reduce $\Phi_{\rm BH}$ by factor $\sim 2$, and thus reduce the jet power (in the form of Poynting flux) by factor $\sim 4$.
Such strong reduction of jet power can be expected to propagate along the jet, with significant and long-ranging consequences for jet structure, magnetic acceleration, and particularly for dissipation.
{\cite{2023ApJ...959L...3D} described such jet modulation in terms of waves propagating along the interface between the jet and the wind (i.e., the jet sheath), calculated ray-traced polarimetric synchrotron images of the jet and demonstrated that the net polarization vector traces a loop in the Stokes parameters $Q/I$ vs. $U/I$ during the flux eruption.}
More recently, \cite{2025ApJ...983...77T} described how variations of the width of such jet images are correlated with variations of $\Phi_{\rm BH}$,
{and \cite{2025PhRvD.112f3013C} reported variations of the jet base opening and tilt angles respectively correlated and anti-correlated with $\Phi_{\rm BH}$.}

In this work, in Section \ref{sec_res}, we analyze the results of a 3D general-relativistic magneto-hydro-dynamic (GRMHD) numerical simulation of MAD accretion onto a high-spin ($a=15/16$) Kerr BH, characterized by very high resolution (effectively $N_r = 5376$ and $N_\theta = N_\phi = 2304$ in log-spherical Kerr-Schild coordinates with static mesh refinement along the polar axes), {sufficient to achieve thin current layers tearing into plasmoids that mediate magnetic reconnection} \citep{2022ApJ...924L..32R}.
This simulation was computed using the H-AMR code \citep{2022ApJS..263...26L}, and its results have been explored in several previous studies \citep{2023ApJ...943L..29H,2023ApJ...959L...3D,2023PhRvR...5d3023Z,2023MNRAS.526.2924J,2024MNRAS.533..254S}.
{Our analysis is limited to a single episode of magnetic flux eruption and subsequent re-accumulation.}
In Section \ref{sec_res_geom}, we measure variations in the structure of relativistic jets to distances of\footnote{The gravitational radius is $r_{\rm g} = GM_{\rm BH}/c^2 \equiv M_{\rm BH}$ with $M_{\rm BH}$ the BH mass.
We also define the gravitational time $t_{\rm g} = r_{\rm g}/c$.}
$\sim 10^3 r_{\rm g}$ following a major magnetic flux eruption.
We show that the jet section of reduced power
develops a strong and distinct geometric distortion, partially stabilized by the jet core made of poloidal magnetic field unaffected by the eruption.
In Section \ref{sec_res_flux}, we analyze variations of magnetic flux along the jet spine.
In Section \ref{sec_res_struct}, we describe the intrinsic structure of the jet spine.
In Section \ref{sec_res_connect}, we compute what fractions of magnetic flux connect the BH horizon and the inner accretion flow with the jet spine, jet sheath, and the torus; and how they vary over a full cycle of eruption and accumulation.
In Section \ref{sec_res_bypass}, we show that a large fraction of magnetic flux ejected from the jet base is re-advected and forms structures that we call magnetic bypasses.
Section \ref{sec_disc} presents the discussion, and Section \ref{sec_conc} the conclusions.
In Appendix \ref{sec_res_poloidal_lines}, we present the intrinsic structure of poloidal field lines and magnetic pitch.
In Appendix \ref{sec_res_impact}, we describe an oblique interface {forming} between the re-powered jet spine and the jet sheath.

We use the natural Heaviside-Lorentz units with the gravitational constant $G = 1$, the speed of light $c = 1$, and the $(4\pi)^{1/2}$ denominator absorbed into the magnetic field $B^i$ and the magnetic 4-vector $b^\mu$.
The Greek indices span $\{0,1,2,3\}$, and the Latin indices span $\{1,2,3\}$.

\section{Results}
\label{sec_res}


{We begin by presenting the context for our analysis of a single BH flux eruption and the resulting jet distortion followed to the distance of $z \sim 850 r_{\rm g}$.}

\begin{figure}[h]
\includegraphics[width=\columnwidth]{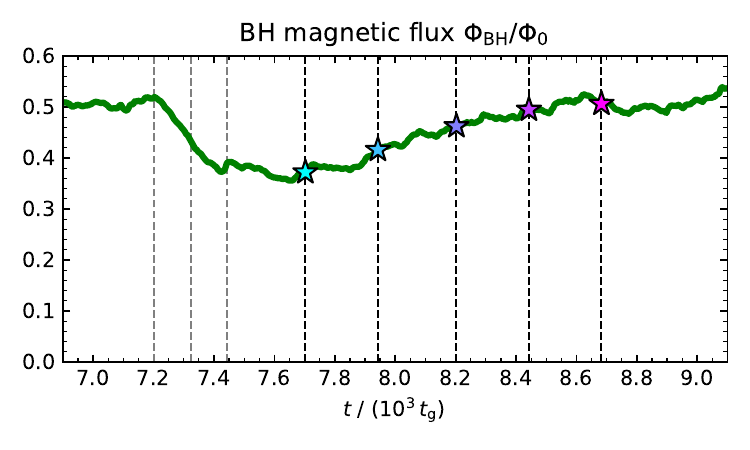}
\caption{Time evolution of unsigned magnetic flux across the BH horizon $\Phi_{\rm BH}$ as fraction of the total torus flux $\Phi_0$.
Epochs corresponding to the $(r,\phi)$-maps shown in Figure \ref{fig_rphmaps} are indicated with the gray vertical dashed lines.
Epochs corresponding to the $(x,z)$-maps shown in Figure \ref{fig_xzmaps_Bz} are indicated with the black vertical dashed lines.
The colored stars correspond to those marked in Figure \ref{fig_flux_spine}.
}
\label{fig_hist}
\end{figure}

Figure \ref{fig_hist} presents the time evolution of $\Phi_{\rm BH}(t)$ as fraction of the total torus flux ($\Phi_0 = 161.11$ in code units) for an episode of the simulation including a magnetic flux eruption (decrease of $\Phi_{\rm BH}$ {by $\simeq 29\%$ over $\Delta t_{\rm erupt} \simeq 230 t_{\rm g}$} from $\simeq 0.52\Phi_0$ at $t = 7.2k t_{\rm g}$ to $\simeq 0.37\Phi_0$ at $t = 7.43k t_{\rm g}$) and subsequent accumulation of magnetic flux on the BH back to $\Phi_{\rm BH} \simeq 0.52\Phi_0$ by $t = 8.63k t_{\rm g}$ {($\Delta t_{\rm accum} \sim 1\,{\rm kt_{\rm g}}$)}.

\begin{figure*}
\includegraphics[width=\textwidth]{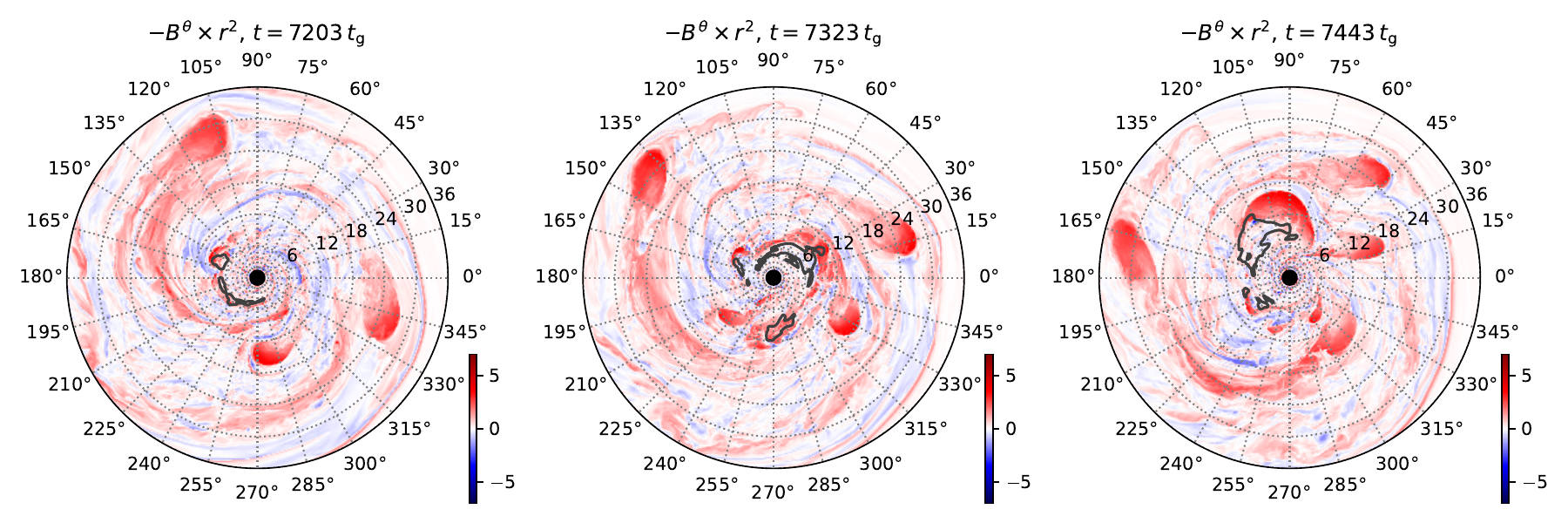}
\caption{Maps in the $(r,\phi)$ coordinates ($r < 36 r_{\rm g}$) of slices along the equatorial plane $\theta = \pi/2$ of `vertical' (on the equator) magnetic field component $-B^\theta$ (multiplied by $r^2$ to compensate for its radial decay) showing magnetic flux tubes (`hotspots'; red).
The 3 panels show different time epochs spanning a magnetic flux eruption, as indicated with the gray vertical dashed lines in Figure \ref{fig_hist}.
The dark-gray contours mark regions of strongly positive radial velocity $v^r > 0.1$.}
\label{fig_rphmaps}
\end{figure*}

Figure \ref{fig_rphmaps} shows {the `vertical' magnetic field component $-B^\theta$ (scaled by $r^2$) in the equatorial plane $\theta = \pi/2$ during 3 stages of} the BH flux eruption.
In the outer regions there are {flux tubes (regular regions of positive $-B^\theta$; also known as hotspots since they contain plasma heated by reconnection during the eruption; \citealt{2020ApJ...900..100R})} ejected during previous eruptions: one of them orbits at $r/r_{\rm g} \simeq 28-32$ counter-clockwise (CCW) from $\phi \simeq 110^\circ$ to $\phi \simeq 160^\circ$; another at $r/r_{\rm g} \simeq 24$ orbits CCW from $\phi \simeq 340^\circ$ to $\phi \simeq 50^\circ$.
The freshly ejected magnetic flux can be identified using ejection regions of positive radial velocity $v^r > 0.1$ marked by dark-gray contours: the flux tubes are located directly downstream (at higher $r$) from those ejections.
{In the middle stage of the eruption ($t \simeq 7.32 kt_{\rm g}$) it extends over $30^\circ \lesssim \phi \lesssim 75^\circ$; by the late stage ($t \simeq 7.44 kt_{\rm g}$) it rotated CCW to $90^\circ \lesssim \phi \lesssim 150^\circ$.}

{In this work we analyze a particular example of BH flux eruption with magnetic flux decreasing by $\sim 30\%$ over $\Delta t_{\rm erupt} \simeq 230 t_{\rm g}$ and accumulating to the initial level over $\Delta t_{\rm accum} \sim 1\,kt_{\rm g}$, ejecting a fresh magnetic flux tube roughly into the $90^\circ \lesssim \phi \lesssim 150^\circ$ sector of the equatorial plane.}

\vskip 2em
\subsection{Geometry of the jet spine}
\label{sec_res_geom}

{Jet spine is the magnetically dominated region of the relativistic jet; following \cite{2018A&A...612A..34D,2024MNRAS.533..254S}, we define it using the cold magnetization\footnote{The cold magnetization is defined as $\sigma_{\rm c} = b^2/\rho$, based on the magnetic 4-vector $b^\mu$ ($b^2 \equiv b_\mu b^\mu$) and the rest-frame plasma mass density $\rho$. The more general hot magnetization $\sigma_{\rm h} = b^2/w$ would use the relativistic enthalpy density $w = \rho + (13/4)P$ (for the adiabatic index of $13/9$) with $P$ the plasma pressure. Since the jet spine-sheath boundary is relativistically cold with temperature $T = P/\rho \ll 1$, $w \simeq \rho$ and $\sigma_{\rm h} \simeq \sigma_{\rm c}$.} as the region where $\sigma_{\rm c} > 1$.  We measure the geometry of jet spine as function of distance $z$, finding a strong distortion evolving over time following the BH flux eruption.}

\begin{figure*}
\includegraphics[width=\textwidth]{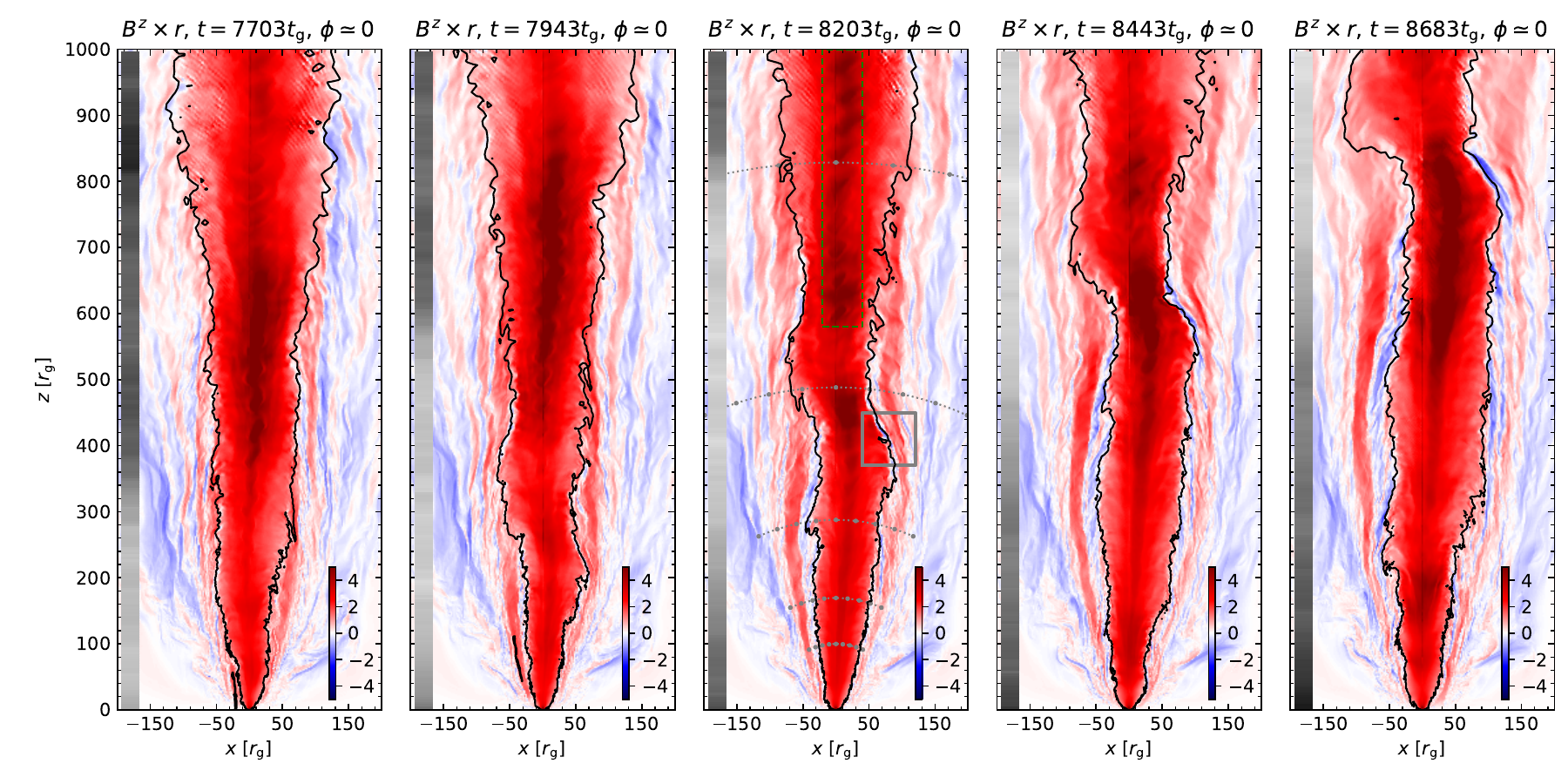}
\caption{Maps in the $(x,z)$ coordinates of slices along the $\phi=0,\pi$ plane of `axial' magnetic field component $B^z$ (multiplied by $r$; red: $B^z > 0$, blue: $B^z < 0$), showing the time evolution of jet spine geometry following a BH flux eruption, delineated by contours of cold magnetization $\sigma_{\rm c} = 1$ (black lines).
The 5 panels show different time epochs spanning the BH flux accumulation phase , as indicated with the black vertical dashed lines in Figure \ref{fig_hist}.
The gray bars on the left side of each panel indicate the magnetic flux through jet spine $\Phi_{\rm sp}(z)$ in shades of gray (white would correspond to 40, black to 85; cf. the upper panel of Figure \ref{fig_flux_spine}).
In the middle panel, the gray dotted arcs represent the grid of $\theta$ angles presented in Figure \ref{fig_thphmaps}, the gray square boxes mark the oblique impact region presented in detail in Figure \ref{fig_xzmaps_shock}, {and the green dashed box marks the region showing possible imprints of the current-driven kink instability}.}
\label{fig_xzmaps_Bz}
\end{figure*}

Figure \ref{fig_xzmaps_Bz} shows a sequence of jet images evolving in time after the BH flux eruption.
The jet spine is indicated by black contours.
Most of the positive axial magnetic field $B^z > 0$ is concentrated within the spine, contributing to its net magnetic flux $\Phi_{\rm sp} > 0$.
Outside the jet spine, one can notice distinct tubes of positive axial field ($B^z > 0$), as well as extensive regions of negative axial field ($B^z < 0$).
The first panel from the left shows a relatively regular jet spine at $t \simeq 7.7k t_{\rm g}$ with little positive axial field outside the spine.
On the left side of the base of this jet, a magnetic flux tube connects to the equatorial plane at $x \simeq -22r_{\rm g}$.
In the second panel, at $t \simeq 7.94k t_{\rm g}$, that flux tube extends to the distance of $z \simeq 270 r_{\rm g}$, where it connects to the spine, which is extended to the left.
Moreover, the entire spine bends to the left over the range of $200 \lesssim z/r_{\rm g} \lesssim 320$ -- we refer to this section as the jet distortion.
In the third panel, at $t \simeq 8.2k t_{\rm g}$, the jet distortion is located at $400 \lesssim z/r_{\rm g} \lesssim 500$, the flux tube on the left side connects likewise.
One can also notice a layer of reversed axial field (blue; $B^z < 0$) between that flux tube and the spine, e.g., around $x \simeq -40 r_{\rm g}$ and $z \sim 330 r_{\rm g}$.
As shown in the 4th and 5th panels, the jet distortion propagated to $580 < z/r_{\rm g} < 680$ at $t \simeq 8.44k t_{\rm g}$, and to $760 < z/r_{\rm g} < 860$ at $t \simeq 8.68k t_{\rm g}$.
It thus propagates linearly with the speed of $\simeq 0.75c$ and maintains a constant thickness of $\Delta z \sim 100 r_{\rm g}$.
On the right side of the jet distortion another region of reversed axial field {(blue: $B^z < 0$)} develops, as can be seen most prominently in the 5th panel (e.g., at $x \simeq 95 r_{\rm g}$ and $z \simeq 820 r_{\rm g}$).
{In Appendix \ref{sec_res_impact} we will show that} in this region the distorted jet impacts the surrounding sheath medium.

\begin{figure}
\includegraphics[width=\columnwidth]{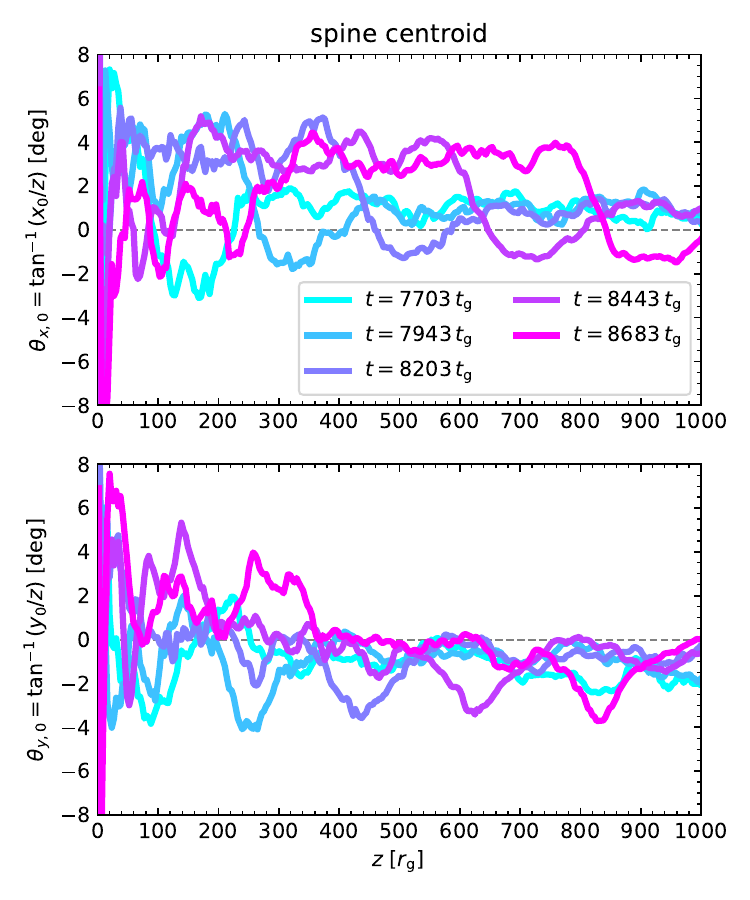}
\caption{Centroids $(x_0,y_0)$ of ellipses fitted to the jet spine boundary ($\sigma = 1$ contour), functions of distance $z$ for the 5 epochs presented in Figure \ref{fig_xzmaps_Bz}.
The panels show angles spanned by the coordinates $x_0$ (upper) and $y_0$ (lower).}
\label{fig_centroid_spine}
\end{figure}

Figure \ref{fig_centroid_spine} presents the angular deflection of jet spine centroid $\theta_{x,0} = \tan^{-1}(x_0/z)$, $\theta_{y,0} = \tan^{-1}(y_0/z)$ calculated by fitting an ellipse to the spine boundary (contour of $\sigma = 1$) along distance $z$.
The $\theta_{x,0}$ angle shows variations between $\simeq -1.25^\circ$ and $\simeq +4^\circ$.
In the first epoch ($t \simeq 7.7k t_{\rm g}$), the spine for $z > 240 r_{\rm g}$ is directed at $\theta_{x,0} \simeq +1^\circ$.
In later epochs this is followed (propagating towards increasing $z$) by a short section deflected to $\theta_{x,0} \simeq -1.25^\circ$, followed by the \emph{main {deflection}} to $\theta_{x,0} \simeq +4^\circ$, the value that is maintained over a long section {of re-powered jet}.
In the last epoch ($t \simeq 8.68k t_{\rm g}$), the re-powered jet section extends over $330 \lesssim z/r_{\rm g} \lesssim 790$.
At the same time, the other angle $\theta_{y,0}$ shows no deflection along the re-powered jet section, but it shows deflection to $\theta_{y,0} \sim -3^\circ$ in the main {deflection} (peaking at $z \simeq 830 r_{\rm g}$), and opposite deflection to $\theta_{y,0} \sim +3^\circ$ in the trailing distortion (peaking at $z \simeq 260 r_{\rm g}$).
{The eruption that produced this jet distortion was directed mainly towards $+y$ ($\phi \sim 90^\circ$; Figure \ref{fig_rphmaps}), roughly opposite to the $-y$ direction of the main {deflection} of the jet.}

{In the aftermath of magnetic flux eruption, the jet spine centroid is thus tilted by up to $\sim 4^\circ$ from the fixed coordinate axis.
The jet spine distortion consists of 2 sections: a helical distortion over $\Delta z \sim 100 r_{\rm g}$ corresponding to the weakest jet, followed by a re-powered jet section with constant deflection over $\Delta z \sim 450 r_{\rm g}$.}

\subsection{Magnetic flux through the jet spine}
\label{sec_res_flux}

{Magnetic flux across the BH horizon $\Phi_{\rm BH}$ varies strongly in time with a characteristic pattern (Figure \ref{fig_hist}).
Here we report the magnetic flux across the jet spine as function of distance along the jet, and compare its variation in time with $\Phi_{\rm BH}(t)$.}

\begin{figure}
\includegraphics[width=\columnwidth]{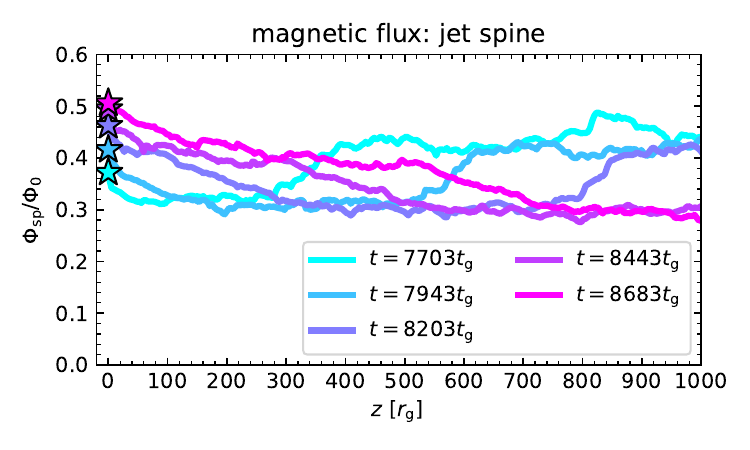}
\includegraphics[width=\columnwidth]{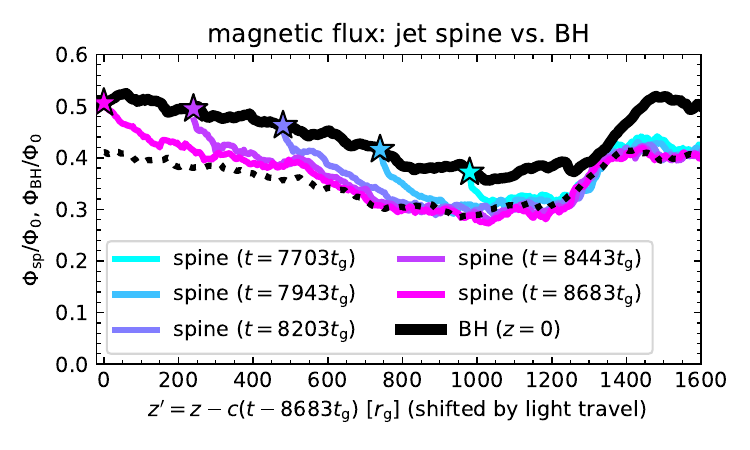}
\caption{Magnetic flux $\Phi_{\rm sp}$ (as fraction of the total torus flux $\Phi_0$) through the jet spine measured at different distances $z$ for the 5 epochs shown in Figure \ref{fig_xzmaps_Bz}.
The color stars indicate the BH flux $\Phi_{\rm BH}/\Phi_0$ for the 5 epochs, also marked in Figure \ref{fig_hist}.
Upper panel: as function of distance $z$.
Lower panel: as function of $z' = z - c(t-t_5)$ (distance shifted by light travel wrt. $t_5 \simeq 8683 t_{\rm g}$), compared with $\Phi_{\rm BH}(t)/\Phi_0$ (solid black line; $z=0$, hence function of $z' = -c(t-t_5)$) (dotted black line: $\Phi_{\rm BH}(z' - 80r_{\rm g})/\Phi_0 \times 0.8$).
}
\label{fig_flux_spine}
\end{figure}

The upper panel of Figure \ref{fig_flux_spine} shows the magnetic flux $\Phi_{\rm sp}$ (as fraction of $\Phi_0$) through the jet spine (defined by $\sigma > 1$) as function of distance $z$ for the 5 epochs presented in Figure \ref{fig_xzmaps_Bz}.
These epochs correspond to the accumulation of magnetic flux at the BH (cf. Figure \ref{fig_hist}), hence for $z < 100$ the profiles of $\Phi_{\rm sp}$ increase over time, closely matching $\Phi_{\rm BH}$ (marked with stars).
For the first epoch ($t \simeq 7.7k t_{\rm g}$), $\Phi_{\rm sp}(z)$ has a flat minimum of $\Phi_{\rm sp} \simeq 0.32\Phi_0$ extending over $40 < z/r_{\rm g} < 280$, then increases over $\Delta z \simeq 110 r_{\rm g}$ to a higher level of $\Phi_{\rm sp} \simeq 0.42\Phi_0$ extending over $390 < z/r_{\rm g} < 750$.
For the next 2 epochs, we find the same pattern shifted towards larger distances; for $t \simeq 8.2k t_{\rm g}$ the minimum level of $\Phi_{\rm sp} \simeq 0.3\Phi_0$ extends over $370 < z/r_{\rm g} < 720$, and a higher level of $\Phi_{\rm sp} \simeq 0.41\Phi_0$ begins at $z \simeq 870 r_{\rm g}$.
What follows the flat minimum at later epochs is a gradual increase of magnetic flux; for $t \simeq 8.69k t_{\rm g}$, it increases from $\Phi_{\rm sp} \simeq 0.3\Phi_0$ at $z \simeq 790 r_{\rm g}$ to $\Phi_{\rm sp} \simeq 0.5\Phi_0$ at $z = 0$.

The lower panel of Figure \ref{fig_flux_spine} shows the same profiles $\Phi_{\rm sp}$ of magnetic flux through jet spine as functions of distance shifted by light-travel $z' = z - c(t-t_5)$ with $t_5 \simeq 8.69k t_{\rm g}$.
Such profiles are well aligned in the sections of flat minima, sharp flux increase with $z$, and flat high-level sections.
One should notice that these sharp increases of spine magnetic flux $\Phi_{\rm sp}$ (propagating with speed $\simeq c$) do not coincide with the geometric distortion of the jet spine (propagating with speed $\simeq 0.75c$),
{although the distortion is within the section of low $\Phi_{\rm sp}$ during those 5 epochs.}
This is not obvious from maps of $B^z$, hence in each panel of Figure \ref{fig_xzmaps_Bz} we indicate the profile of $\Phi_{\rm sp}(z)$ with a gray strip of variable shading.
The spine regions with most intense axial field ($B^z \sim 5/r$ in code units) are not good indicators of high flux $\Phi_{\rm sp}$, which is also sensitive to variations of the spine cross section.

Such shifted profiles of $\Phi_{\rm sp}(z')$ can also be compared with the time history of BH magnetic flux $\Phi_{\rm BH}(t)$, with time running from right to left.
The sharp increases of $\Phi_{\rm sp}$ at $z' \sim (1200-1400) r_{\rm g}$ match the profile of $\Phi_{\rm BH}(t)$ during the magnetic flux eruption with a small shift of $\Delta z' \sim 80 r_{\rm g}$ (see the dotted line in the lower panel of Figure \ref{fig_flux_spine}; this is significantly more than the gravitational light-travel delay of $\Delta t_{\rm gltd} \sim 2t_{\rm g}\ln(z/r_{\rm g}) \simeq 14t_{\rm g}$ between the BH horizon and $z \sim 10^3 r_{\rm g}$, only weakly dependent on the BH spin).
The overall level of jet spine magnetic flux is $\sim 20\%$ lower than $\Phi_{\rm BH}(t)$.
Analysis of the sample of lines integrated from the BH horizon (Section \ref{sec_res_connect}) confirms that lines representing such a fraction of BH flux cross the spine/sheath boundary within $z \lesssim 100\,r_{\rm g}$.

{We thus find that the magnetic flux profiles along the jet spine $\Phi_{\rm sp}(z-c(t-t_0))$ closely reflect the temporal evolution of the BH magnetic flux $\Phi_{\rm BH}(t)$.}

\subsection{Intrinsic structure of the jet spine}
\label{sec_res_struct}

{Since the jet spine is geometrically distorted, its physical axis is shifted from the fixed coordinate axis. Here we show that in local coordinates centered at the physical axis at the spheres of different radii $r$, both the shape of the jet spine and its various parameters maintain a high level of axial symmetry.}

\begin{figure}
\includegraphics[width=0.495\columnwidth]{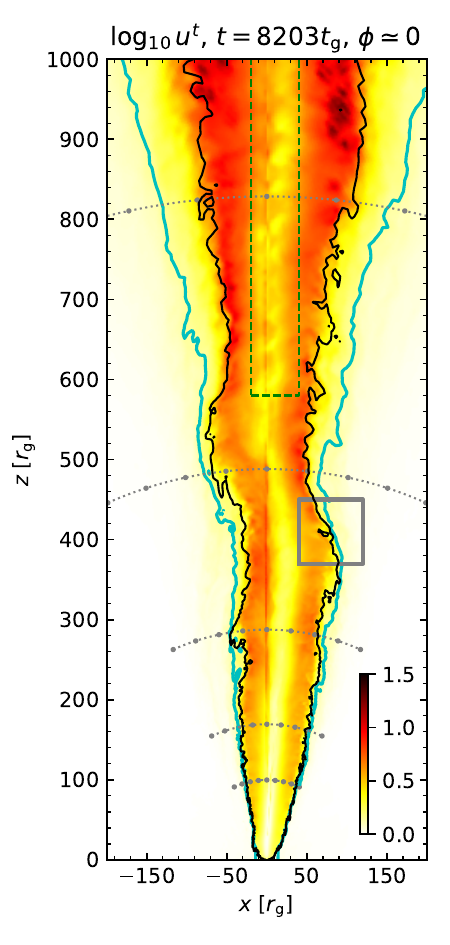}
\includegraphics[width=0.495\columnwidth]{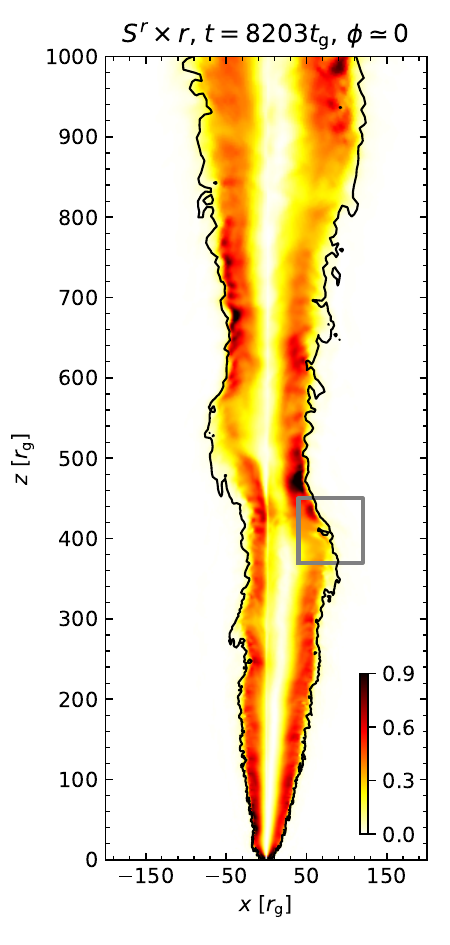}
\caption{Maps in the $(x,z)$ coordinates of slices along the $\phi=0,\pi$ plane for the time epoch $t \simeq 8.2k t_{\rm g}$ corresponding to the middle panel of Figure \ref{fig_xzmaps_Bz}.
The left panel shows the Lorentz factor $u^t$ in log color scale, the right panel shows the Poynting flux $S^r$ (multiplied by $r$).
The black lines mark the jet spine boundary (contour of cold magnetization $\sigma_{\rm c} = b^2/\rho = 1$).
The gray square box marks the oblique impact region presented in detail in Figure \ref{fig_xzmaps_shock}.
{In the left panel,
the cyan solid lines mark the contour of $u^t = \sqrt{2}$ outside the jet spine (the jet sheath boundary),
the gray dotted arcs represent the grid of $\theta$ angles presented in Figure \ref{fig_thphmaps},
and the green dashed box marks the region showing possible imprints of the current-driven kink instability (cf. the middle panel of Figure \ref{fig_xzmaps_Bz}).}}
\label{fig_xzmaps2}
\end{figure}

The left panel of Figure \ref{fig_xzmaps2} shows the $(x,z)$ map of bulk Lorentz factor $u^t$.
Relativistic motion is largely confined to the jet spine, but it also extends to the surrounding layer referred to as the \emph{jet sheath}, the boundary of which is {chosen} as the contour of Lorentz factor $u^t = \sqrt{2}$ (corresponding to the 4-velocity $[(u^t)^2-1]^{1/2} = 1$).
Within the jet spine, $u^t$ shows a strong lateral structure, increasing from the center to the spine boundaries.
This is consistent with the axisymmetric models of relativistic jet acceleration, which predict inefficient acceleration along the {jet core (the central region of the jet spine where magnetic field is dominated by the poloidal component)} and the inner profile of Lorentz factor proportional to cylindrical radius \citep[e.g.,][]{2009MNRAS.397.1486B}.

The right panel of Figure \ref{fig_xzmaps2} shows the $(x,z)$ map of radial Poynting flux $S^r$ scaled by $r$.
Instead of the GR {radial Poynting flux density (electromagnetic momentum tensor)}
\be
S^r = T_{\rm EM}^{0r} = b^2 u^0 u^r - b^0 b^r - \frac{b^2}{2} g^{0r}
\ee
(with $b^\mu$ the magnetic 4-vector, $u^\mu$ the 4-velocity, $g^{\mu\nu}$ the inverse metric tensor),
we plot the flat-spacetime approximation
\be
S^r \simeq B^2 v^r - (v_iB^i)B^r\,,
\label{eq_Sr_Mink1}
\ee
the difference between them is negligible at distances $r > 10 r_{\rm g}$.
The Poynting flux is well contained within the $\sigma = 1$ contour (jet spine), flowing through most of the available cross section away from the jet core, including the distortion at $z \sim (400-500) r_{\rm g}$.

\begin{figure}
\includegraphics[width=0.495\columnwidth]{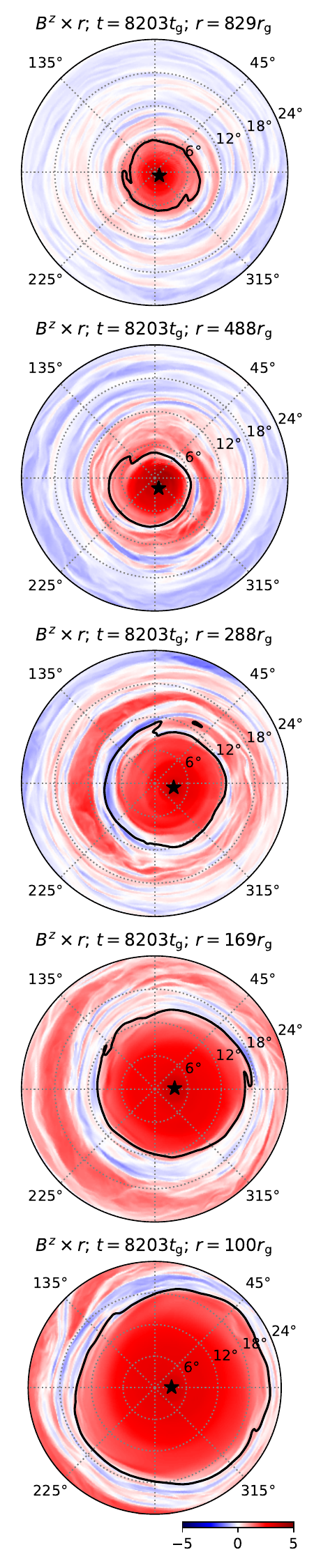}
\includegraphics[width=0.495\columnwidth]{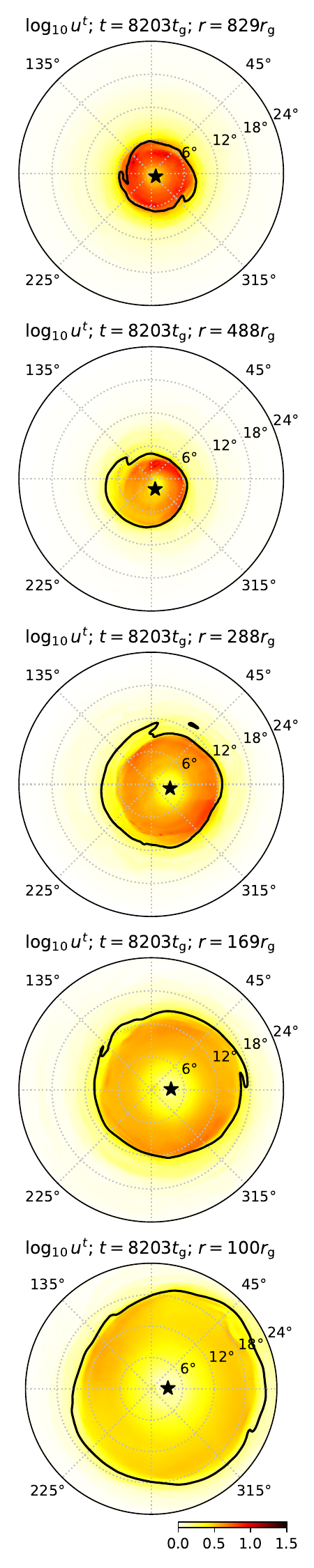}
\caption{Maps in the $\theta \in [0:24^\circ]$ (polar) and $\phi \in [0:2\pi]$ (azimuthal) coordinates for several values of spherical radius $r$ (increasing from bottom to top) of the axial magnetic field $B^z$ (left panels) and of the Lorentz factor $u^t$ (right panels; log color scale).
The black lines indicate the jet spine boundary (contour of cold magnetization $\sigma_{\rm c} = b^2/\rho = 1$).
The black stars mark the position of the jet axis.
}
\label{fig_thphmaps}
\end{figure}

Figure \ref{fig_thphmaps} shows maps of axial magnetic field $B^z$ and bulk Lorentz factor $u^t$ in the $(\theta,\phi)$ coordinates for several values of the spherical radius $r$.
The spine boundary (contour of $\sigma=1$) is indicated with solid black lines.
Maps of $B^z(\theta,\phi)$ show positive magnetic flux concentrated within the jet spine, but also alternating rings of positive and negative $B^z$ outside the spine.
Comparing these maps with the middle panel of Figure \ref{fig_xzmaps_Bz}, one can notice that the ejected flux tubes surround the jet spine in the form of concentric spirals.
The 2nd and 3rd panels (from the top) of Figure \ref{fig_thphmaps} show how such ejected flux tubes connect with the jet spine: for $r \simeq 488 r_{\rm g}$ this happens at azimuth $\phi \simeq 135^\circ$; for $r \simeq 288 r_{\rm g}$ at $\phi \simeq 90^\circ$.
No such connection can be seen in the 4th and 5th panels (for $r/r_{\rm g} \simeq 169, 100$) -- in those cases the jet spine is surrounded at all sides by a layer of reversed axial field $B^z < 0$.
Maps of $u^t(\theta,\phi)$ show regular rings of relativistic motion confined to the less regular jet spine boundary.
The flux connection points in the 2nd and 3rd panels extend CCW into pockets of reduced Lorentz factor.

We use the lateral structure of $u^t$ to determine the position of the {physical} jet axis at given spherical radius $r$.
Consider that point $\bm{P}_0 = (\theta_0,\phi_0)$ is the center of a new coordinate system ($\zeta,\psi$) on the sphere of $r = r_0$ (hence $\zeta_0 = 0$ at $\bm{P}_0$).
For any other point $\bm{P}_1 = (\theta_1,\phi_1)$, we need to calculate its coordinates $(\zeta_1,\psi_1)$.
The angle $\zeta_1$ is the angular distance between $\bm{P}_0$ and $\bm{P}_1$ measured from the origin of spherical coordinates ($r=0$), from the spherical cosines law:
\be
\cos\zeta_1 = \cos\theta_0\cos\theta_1 + \sin\theta_0\sin\theta_1\cos(\phi_0-\phi_1)\,.
\ee
The angle $\psi_1$ is the angle between two great arcs crossing at $\bm{P}_0$: one connecting it to $\bm{P}_1$ (with tangent unit vector $\bm{n}_{01} = \bm{P}_1/\sin\zeta_1-\bm{P}_0/\tan\zeta_1$), the other connecting it to the center $\bm{O}$ of the $(\theta,\phi)$ coordinates on the $r=r_0$ sphere ($\theta=0$ at $\bm{O}$; with tangent unit vector $\bm{n}_{0O} = \bm{O}/\sin\theta_0-\bm{P}/\tan\theta_0$), such that $\cos\psi_1 = \bm{n}_{01}\cdot\bm{n}_{0O}$.
For any radius $r_0$ we determine the position $(\theta_0,\phi_0)$ of the jet axis that minimizes the root-mean-square of $u^t(\zeta)$ profiles over all $\psi$ angles.
In each panel of Figure \ref{fig_thphmaps}, the position of the jet axis is marked with the black star.

\begin{figure}
\includegraphics[width=\columnwidth]{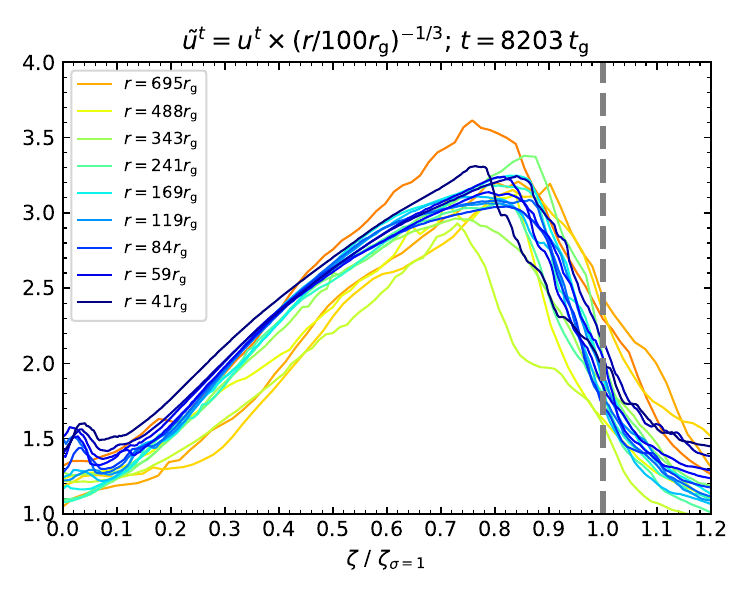}
\caption{Intrinsic profiles of the bulk Lorentz factor, scaled as $\tilde{u}^t = u^t \times (r/100 r_{\rm g})^{-1/3}$, functions of the polar angle $\zeta$ normalized to the spine boundary $\zeta_{\sigma=1}$ (gray dashed line), compared for many values of the spherical radius $r$.}
\label{fig_zetaprofs_ut}
\end{figure}

\begin{figure}
\includegraphics[width=\columnwidth]{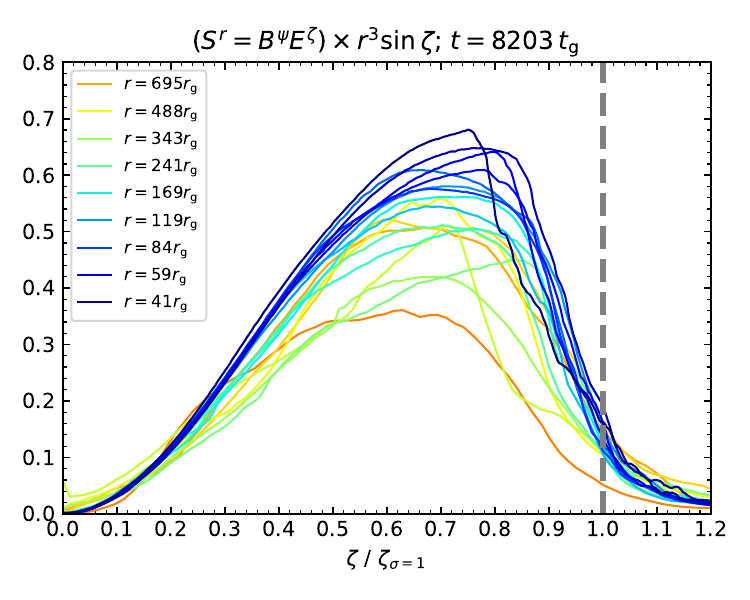}
\caption{Intrinsic profiles of the Poynting flux density $S^r$, functions of the polar angle $\zeta$ normalized to the spine boundary $\zeta(\sigma=1)$ (gray dashed line), compared for many values of the spherical radius $r$.}
\label{fig_zetaprofs_Sr}
\end{figure}

Figure \ref{fig_zetaprofs_ut} compares the intrinsic profiles of bulk Lorentz factor, scaled as $\tilde{u}^t = u^t \times (r/100 r_{\rm g})^{-1/3}$, as functions of polar angle $\zeta$ normalized to the spine boundary $\zeta_{\sigma=1}$ over a broad range of spherical radii $40 < r/r_{\rm g} < 830$.
These profiles appear self-similar, peaking at $\zeta_{\rm peak} \sim 0.8\zeta_{\sigma=1}$ at the value of $\tilde{u}^t_{\rm peak}(r) \simeq 3.1$.
The values at the jet spine boundary ($\zeta=\zeta_{\sigma=1}$; gray dashed line) are roughly $\tilde{u}^t_{\sigma=1} \simeq 2$.
The inner parts of the $\tilde{u}^t$ profiles are roughly linear in $\zeta$, indicating a broad jet core.

Figure \ref{fig_zetaprofs_Sr} compares the intrinsic profiles of the spherical-radial Poynting flux $S^r$ as functions of $\zeta/\zeta_{\sigma=1}$.
Here we plot a flat-spacetime approximation in the $(r,\zeta,\psi)$ coordinates, further simplified from Eq. (\ref{eq_Sr_Mink1}) due to axial symmetry ($\psi$-averaged $B^\zeta=0$):
\be
S^r \simeq B^\psi E^\zeta = B^\psi (B^\psi v^r - B^r v^\psi)\,.
\ee
Such Poynting flux is very well confined to the jet spine, peaking at $\zeta/\zeta_{\sigma=1} \simeq 0.7$, with $\sim (90-95)\%$ of cumulative Poynting flux
\vspace{-1ex}
\be
\mathcal{S}^r(\zeta) = 2\pi r^2\int_0^\zeta {\rm d}\zeta'\,\zeta'\,S^r(\zeta')
\ee
within the jet spine.

{We have used distributions of the Lorentz factor $u^t$ in $(\theta,\phi)$ to determine the physical jet axis, introduced shifted polar coordinates $(\zeta,\psi)$, and presented $\psi$-averaged profiles of $u^t(\zeta)$ and Poynting flux $S^r(\zeta)$, determining that $S^r$ is very well contained within the jet spine.}

\subsection{Magnetic connections}
\label{sec_res_connect}

For every analyzed simulation time, we integrated two large samples of magnetic field lines.
The first sample -- \emph{the BH lines} -- was integrated from initial positions just outside the BH horizon ($r_0 \simeq 1.348 r_{\rm g}$) over the entire upper hemisphere ($\theta_0 < 90^\circ$, $0 < \phi_0 < 360^\circ$) where radial magnetic field points outwards ($B^r > 0$).
The second sample -- \emph{the equatorial lines} -- was integrated from initial positions just above the equatorial plane ($\theta_0 \simeq 89.88^\circ$) over the inner accretion flow ($r_{\rm H} < r_0 < r_{\rm br}$, $0 < \phi_0 < 360^\circ$) where the `vertical' magnetic field points to the upper hemisphere ($B^z \equiv -r B^\theta > 0$).
The outer radius of $r_{\rm br} = 32 r_{\rm g}$ corresponds to a break in the cumulative magnetic flux through the equatorial plane
\be
\Phi_{\rm eq}(r) = \int_{r_{\rm in}}^r {\rm d}\tilde{r}\int_0^{2\pi}{\rm d}\phi\;[\sqrt{-g} B^\theta](\tilde{r},\theta_0)
\ee
(for $\theta_0 \simeq 89.88^\circ$ and $r_{\rm in} = r_{\rm H}$),
coinciding with the outer range of the magnetic flux tubes (cf. Figure \ref{fig_rphmaps}).
Each magnetic line was integrated using a 4th order Runge-Kutta scheme with integration step proportional to the local radial cell size $\Delta s = 0.03\Delta r$ until it reached one of the three boundaries: inner radial (BH horizon) $r_{\rm in} = r_{\rm H}$, outer radial $r_{\rm out} = 1700 r_{\rm g}$, or equatorial $\theta_{\rm eq} = 90^\circ$.
The integration was performed in a sub-sampled grid of resolution $N_r/M \times N_\theta/M \times N_\phi/M$ with $M=2$ (convergence was checked against a sparser sub-grid for $M=6$).
The samples of BH lines contain up to $\sim 6.6\times 10^5$ lines,
and the samples of equatorial lines up to $\sim 1.3\times 10^6$ lines.

\begin{figure*}
\includegraphics[width=0.495\textwidth]{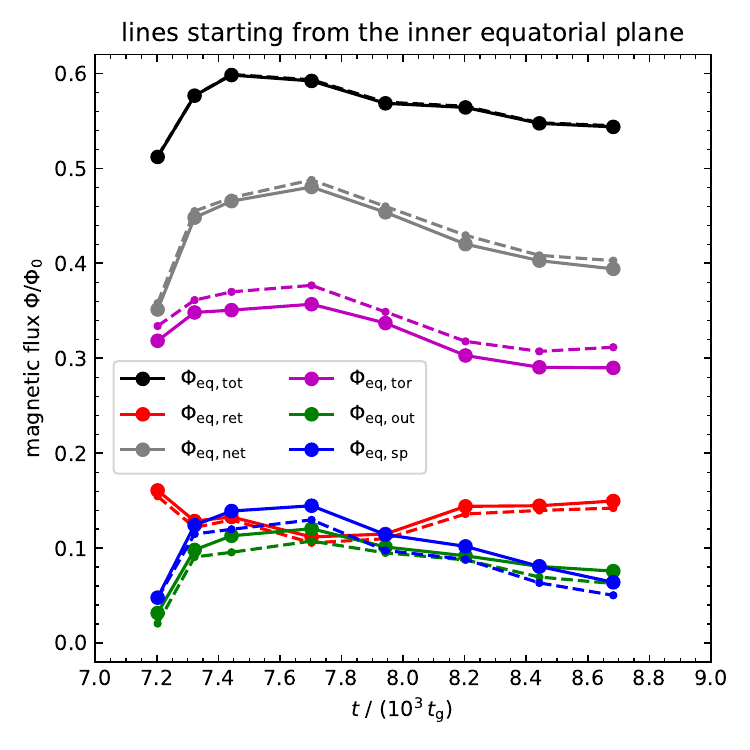}
\includegraphics[width=0.495\textwidth]{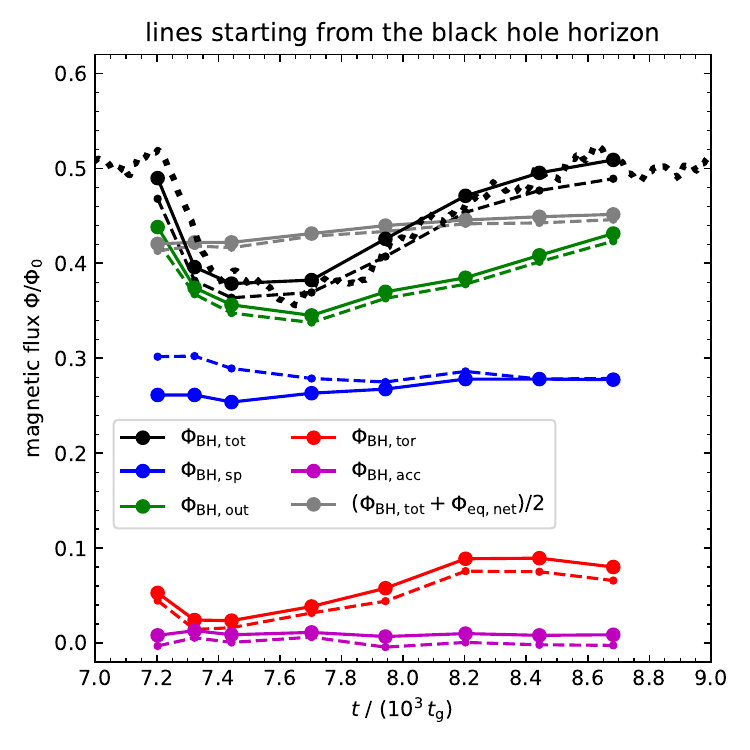}
\caption{Magnetic flux $\Phi$ as fraction of the total torus flux $\Phi_0$ divided according to the magnetic connection of individual field lines, compared for several simulation times.
Left panel: fluxes integrated over the line samples seeded along the equatorial plane for $r_{\rm H} < r < r_{\rm br} = 32 r_{\rm g}$ where $B^z > 0$:
$\Phi_{\rm eq,tot}$ -- total flux;
$\Phi_{\rm eq,ret}$ -- flux returning to the inner equatorial plane;
$\Phi_{\rm eq,net}$ -- net flux leaving the inner accretion flow;
$\Phi_{\rm eq,out}$ -- flux reaching the outer boundary;
$\Phi_{\rm eq,sp}$ -- maximum flux within the jet spine;
$\Phi_{\rm eq,tor}$ -- flux reaching the outer torus.
Right panel: fluxes integrated over the line samples seeded along the BH horizon in the upper hemisphere where $B^r > 0$:
$\Phi_{\rm BH,tot}$ -- total flux;
$\Phi_{\rm BH,out}$ -- flux reaching the outer boundary;
$\Phi_{\rm BH,sp}$ -- flux never leaving the jet spine;
$\Phi_{\rm BH,tor}$ -- flux reaching the outer torus;
$\Phi_{\rm BH,acc}$ -- flux reaching the inner accretion flow.}
\label{fig_flux_division}
\end{figure*}

Figure \ref{fig_flux_division} presents the time evolution of magnetic flux components during a BH flux eruption ($7.2k \lesssim t/t_{\rm g} \lesssim 7.45k$) followed by a BH flux accumulation ($7.7k \lesssim t/t_{\rm g} \lesssim 8.7k$).
The magnetic flux components were integrated over the 2 line samples, and subdivided according to the magnetic connection.

Among the sample of equatorial lines we distinguish:
the total flux pointing upwards (where $B^z > 0$) $\Phi_{\rm eq,tot}$,
the flux returning to the inner equatorial plane $\Phi_{\rm eq,ret}$ (with $B^z < 0$),
the net flux leaving the inner equatorial plane $\Phi_{\rm eq,net} = \Phi_{\rm eq,tot} - \Phi_{\rm eq,ret}$,
the net flux reaching the outer torus ($z \simeq 0$, $r > r_{\rm peak} \simeq 185 r_{\rm g}$) $\Phi_{\rm eq,tor}$,
the net flux reaching the outer boundary ($z > 1700 r_{\rm g}$) $\Phi_{\rm eq,out}$,
and the maximum net flux within the jet spine $\Phi_{\rm eq,sp}$ {(specifically this means $\max_r\{\Phi_{\rm eq,\sigma>1}(r)\}$)}.

Among the sample of BH lines we distinguish:
the total flux pointing outwards (where $B^r > 0$) $\Phi_{\rm BH,tot}$,
the flux reaching the outer boundary ($z > 1700 r_{\rm g}$) $\Phi_{\rm BH,out}$,
the flux never leaving the jet spine $\Phi_{\rm BH,sp}$,
the flux reaching the outer torus ($z \simeq 0$, $r > r_{\rm peak}$) $\Phi_{\rm BH,tor}$.
Magnetic connection between BH horizon and the inner accretion flow is negligible, and $\Phi_{\rm BH,tot} \simeq \Phi_{\rm BH,out} + \Phi_{\rm BH,tor}$.

The total BH flux $\Phi_{\rm BH,tot}$ decreased during the BH flux eruption ($7.2k \lesssim t/t_{\rm g} \lesssim 7.45k$) $\simeq 0.49 \Phi_0$ to $\simeq 0.38\Phi_0$, then increased during the BH flux accumulation ($7.7k < t/t_{\rm g} < 8.7k$) to $\simeq 0.51\Phi_0$.
At the same time, the net flux across the inner equatorial plane ($r < r_{\rm br}$) $\Phi_{\rm eq,net}$ first increased from $\simeq 0.35\Phi_0$ to $\simeq 0.465\Phi_0$, then decreased to $\simeq 0.4\Phi_0$.
The combined flux $\Phi_{\rm BH,tot}+\Phi_{\rm eq,net}$ shows a slow steady increase (a combination of $\Phi_{\rm BH,tot}+\Phi_{\rm eq,tot}$ is not that steady), indicating that additional flux {was} transferred to the accretion region from the inner torus ($r_{\rm br} < r < r_{\rm peak}$) {during the eruption-accumulation cycle}.

Most of the net equatorial flux $\Phi_{\rm eq,net}$ is connected with the outer torus, during the BH flux accumulation $\Phi_{\rm eq,tor}$ first increased from $\simeq 0.32\Phi_0$ to $\simeq 0.35\Phi_0$, then decreased to $\simeq 0.29\Phi_0$.
The flux connected with the outer boundary $\Phi_{\rm eq,out}$ first increased from $\simeq 0.03\Phi_0$ to $\simeq 0.12\Phi_0$, then decreased to $\simeq 0.075\Phi_0$.
Most of that flux connects with (or at least passes through) the jet spine, $\Phi_{\rm eq,sp} \simeq \Phi_{\rm eq,out}$.
The equatorial flux returning to the inner equatorial plane $\Phi_{\rm eq,ret}$ (excluded from the net flux $\Phi_{\rm eq,net}$) is also substantial, it reaches up to $\simeq 0.16\Phi_0$.

Of the flux accumulated on the BH $\Phi_{\rm BH,tot}$, most is connected with the outer boundary -- from $\simeq 0.35\Phi_0 \sim 90\% \Phi_{\rm BH,tot}$ in the beginning of the flux accumulation phase to $\simeq 0.43\Phi_0 \sim 85\% \Phi_{\rm BH,tot}$ in its end.
Notably, the flux that never leaves the jet spine is very stable throughout the eruption/accumulation cycle at $\Phi_{\rm BH,sp} \simeq 0.27\Phi_0$.

{Over the cycle of magnetic flux eruption and accumulation, magnetic flux shifts between the inner equatorial plane ($r_{\rm br} \simeq 32 r_{\rm g}$ being the effective range of ejected flux tubes) and the BH horizon.
More than half of the BH flux stays within the jet spine, unaffected by the eruption cycle.}

\subsection{Magnetic bypasses}
\label{sec_res_bypass}

A magnetic flux eruption ejects from the jet base a significant fraction of magnetic flux that otherwise would be continuously wound up by the spinning spacetime in force-free condition and carry large Poynting flux along the jet.
Ejected magnetic flux tubes rotate more slowly, attain gently helical geometries, and are loaded by plasma, which decreases their magnetization.
Parts of the magnetic flux tubes are eventually re-advected onto the BH by the accretion flow, we refer to this as the magnetic recycling.
The resulting magnetic field lines can be expected to have the following structure propagating along the jet: (1) a tightly wound (toroidal-dominated; $|B^\phi| > |B^z|$) pre-eruption section at large distance, (2) a loosely wound (poloidal-dominated; $|B^z| > |B^\phi|$) section at intermediate distance that we call a bypass, (3) a tightly wound post-eruption section at short distance.

\begin{figure}
\includegraphics[width=\columnwidth]{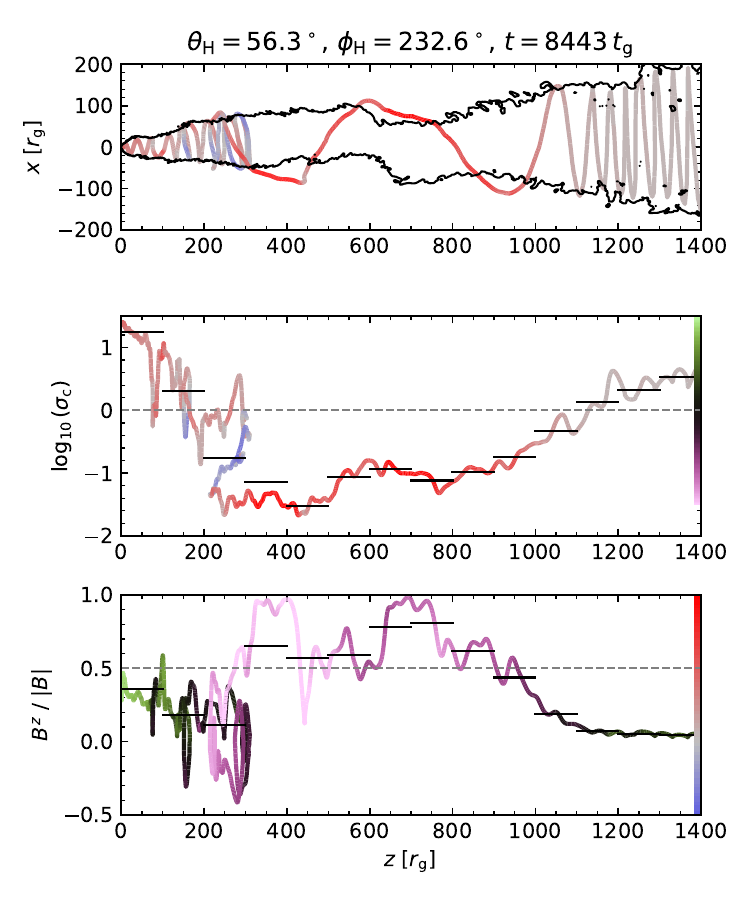}
\caption{An example of magnetic field line with a magnetic bypass.
Integration of this line started just above the BH horizon at the indicated $(\theta_{\rm H},\phi_{\rm H})$.
The upper panel shows the line in the $(x,z)$ space, colored by the local poloidality $B^z/|B|$ (colorbar in the lower panel), and the contour of $\sigma_{\rm c} = 1$ in the $y=0$ plane.
The middle panel shows the local magnetization $\log_{10}(\sigma_{\rm c})$ along the line vs $z$, colored by $B^z/|B|$ (colorbar in the lower panel).
The lower panel shows the local poloidality $B^z/|B|$ along the line vs $z$, colored by $\log_{10}(\sigma_{\rm c})$ (colorbar in the middle panel).
The black lines in the middle and lower panels mark the average values of $\log_{10}(\sigma_{\rm c})$ and $B^z/|B|$, respectively, over line sections with $\Delta z = 100 r_{\rm g}$.}
\label{fig_bypass_example}
\end{figure}

\begin{figure}
\includegraphics[width=\columnwidth]{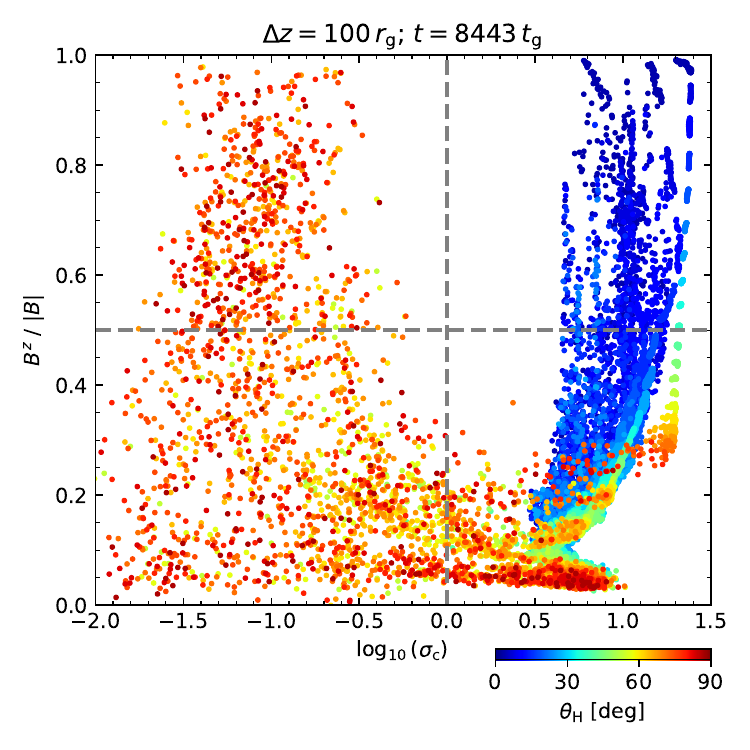}
\caption{Distribution of magnetic poloidality ${\left<B^z / |B|\right>}_z$ vs. magnetization ${\left<\log_{10}\sigma_{\rm c}\right>}_z$ averaged over line sections spanning $\Delta{z} = 100\,r_{\rm g}$ for a complete sample of lines integrated from the BH horizon starting at different angular positions ($\theta_0,\phi_0$). The points are colored by $\theta_0$.}
\label{fig_bypass_sec_ave}
\end{figure}

We have analyzed the sample of BH lines (starting just beyond $r_{\rm H}$ over $0 < \theta_{\rm H} < 90^\circ$ and $0 < \phi_{\rm H} < 360^\circ$) according to the \emph{poloidality} parameter $B^z/|B|$ (with $|B^z/B| < 1/2$ indicating domination of the toroidal component, and $|B^z/B| > 1/2$ -- domination of the poloidal component), and cold magnetization $\sigma_{\rm c} = b^2/\rho$.
We selected only the lines that reach an outer boundary $z_{\rm out} = 1800 r_{\rm g}$.
Each line was divided into sections of equal $\Delta z = 100 r_{\rm g}$, and within each such section the average values of poloidality ${\left<B^z/|B|\right>}_z$ and log-magnetization ${\left<\log_{10}\sigma_{\rm c}\right>}_z$ were calculated (an example is shown in Figure \ref{fig_bypass_example}).
The parameter space of ${\left< B^z/|B|\right>}_z$ vs. ${\left<\log_{10}\sigma_{\rm c}\right>}_z$ (Figure \ref{fig_bypass_sec_ave}) shows two separate concentrations of poloidal-dominated line sections with ${\left< B^z/|B| \right>}_z > 1/2$: highly magnetized (${\left< \log_{10}\sigma_{\rm c} \right>}_z > 0$) and weakly magnetized (${\left< \log_{10}\sigma_{\rm c} \right>}_z < 0$).
The highly magnetized poloidal-dominated sections belong to regular lines along the jet core, anchored to the BH horizon at small polar angles $\theta_{\rm H} < 45^\circ$.
Those lines do not leave the jet spine (have $\sigma_{\rm c} > 1$ over their entire lengths) and are unaffected by the BH flux eruption.
On the other hand, the weakly magnetized poloidal-dominated sections -- the magnetic bypasses -- belong to the lines affected by the BH flux eruption, anchored to the BH at large polar angles $\theta_{\rm H} > 45^\circ$.

\begin{figure*}
\includegraphics[width=\textwidth]{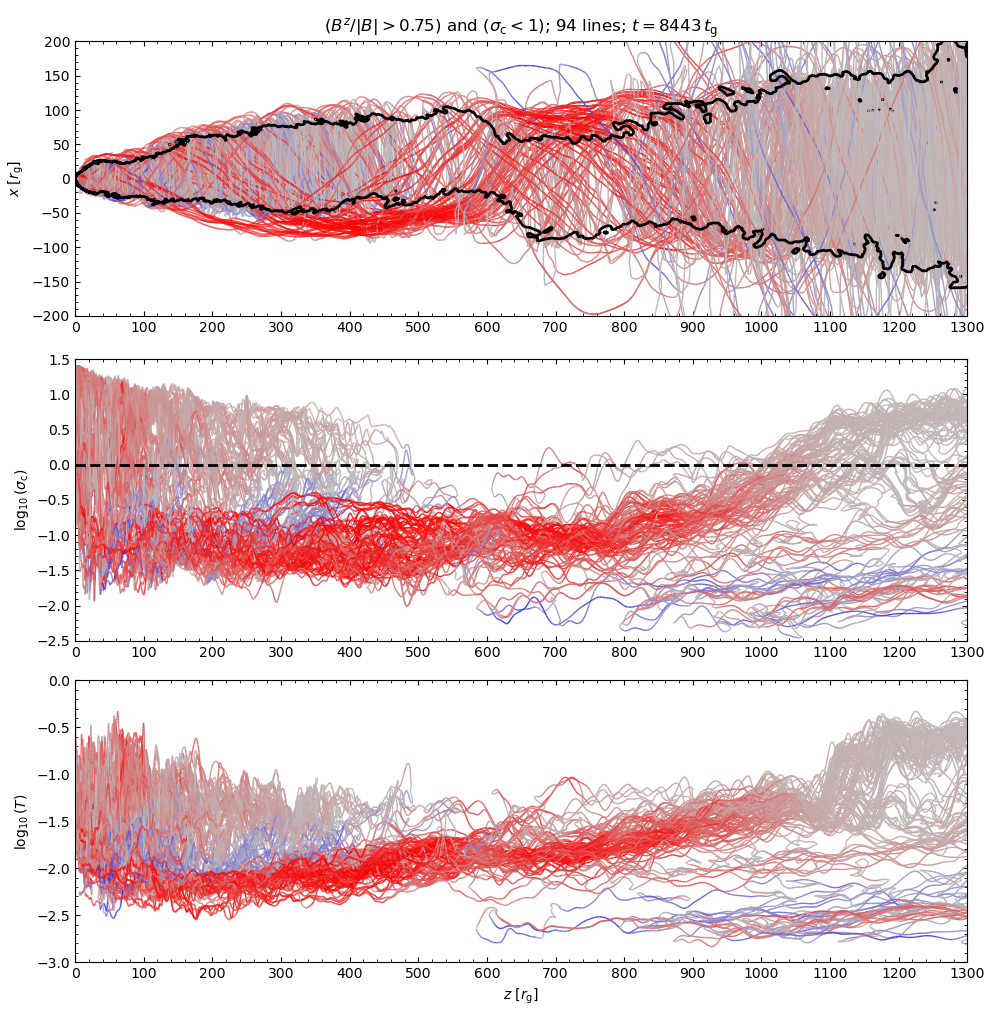}
\caption{A complete sample of magnetic field lines with magnetic bypasses having at least one section with ${\left<B^z/|B|\right>}_z > 3/4$ and ${\left<\log_{10}\sigma_{\rm c}\right>}_z < 0$.
The upper panel shows the line positions in the $(x,z)$ space, and the contour of $\sigma_{\rm c} = 1$ in the $y=0$ plane (black).
The {middle} panel shows the local magnetization $\log_{10}(\sigma_{\rm c})$ along the lines vs $z$.
{The lower panel shows the local plasma temperature $\log_{10}(T)$ along the lines vs $z$.}
In {all} panels the lines are colored by the local poloidality $B^z/|B|$, using the same color scale as in the upper and middle panels of Figure \ref{fig_bypass_example}.}
\label{fig_bypass_sample}
\end{figure*}

Figure \ref{fig_bypass_sample} presents a complete sample of 94 lines (out of a set of 1152 starting positions along the BH horizon at uniform $0 < \theta_{\rm H} < 90^\circ$ and $0 < \phi_{\rm H} < 360^\circ$) having strong bypasses (at least one $\Delta z$ section with ${\left<B^z/|B|\right>}_z > 3/4$ and ${\left<\log_{10}\sigma_{\rm c}\right>}_z < 0$), carrying a magnetic flux of $\simeq 0.037\Phi_0$, which represents $\sim 10\%$ of $\Phi_{\rm BH,out}$ (the BH flux reaching the outer boundary, see Figure \ref{fig_flux_division}) and $\sim 1/3$ of $\Delta\Phi_{\rm BH}$ shed from the BH during the preceding eruption at $t \sim (7.2-7.45) kt_{\rm g}$.
The bypasses are roughly aligned and collectively form a distinct band gently wrapped around the distorted jet spine.
{We also found that plasma temperature along the bypass lines is non-relativistic with $T = P/\rho \sim 10^{-2}$, which is comparable to the temperature of the jet sheath away from the spine-sheath boundary.}

Looking at other stages along the BH flux eruption and accumulation cycle, no magnetic bypasses are found within $z < 1800 r_{\rm g}$ during the BH flux eruption.
The first bypasses appear at the beginning of the BH flux accumulation phase, already being characterized by $\sigma < 1$.
We do not find any poloidal-dominated section of a line anchored at $\theta_{\rm H} > 45^\circ$ with $\sigma > 1$, indicating that \emph{all ejected magnetic flux tubes are loaded with plasma before they are re-advected onto the BH horizon}.

\clearpage
\section{Discussion}
\label{sec_disc}

\subsection{Perturbations in relativistic jets}

Astrophysical relativistic jets are powerful dissipative outflows.
In different systems they need to operate on different time scales, from transient GRBs ($t_{\rm obs} \sim 1\,{\rm s} \sim 2\times 10^5\,r_{\rm g}/c$ for $M_{\rm BH} \sim M_\odot$) to Mpc-scale radio galaxies ($t_{\rm obs} \sim 10^7\,{\rm yr} \sim 6\times 10^{10}\,r_{\rm g}/c$ for $M_{\rm BH} \sim 10^9 M_\odot$).
In many AGN jets discrete bright knots are observed on kpc scales \citep{1994AJ....108..766B} and on pc scales \citep{2009AJ....137.3718L}, at small viewing angles (especially in microquasars and blazars) propagating with apparently superluminal speeds \citep{1999ARA&A..37..409M,2001ApJS..134..181J,2017ApJ...846...98J,2022ApJS..260...12W}.
The kinematics of radio knots is complex, they propagate not only along the jet, but often show perpendicular components of apparent velocity \citep{2016AJ....152...12L} (particularly intriguing is the case of 3C~279; \citealt{2020A&A...640A..69K,2023NatAs...7.1359F}).
{The inner jet of M87 (at least up to $12\;{\rm mas}$) shows {structural variations \citep{2024A&A...692A.140A}}, recently described as oscillations on time scales of $\sim 1\;{\rm yr}$ propagating with apparently superluminal speed $v_{\rm app} \simeq 2.8 c$ \citep{2026ApJ...999..169R}.}

The emergence of jet knots observed in the radio has been linked with changes in the accretion state observed in the X-rays, both in the microquasars and AGNs \citep{1998A&A...330L...9M,2002Natur.417..625M}.
The prevailing scenario for the production of jet knots is the polarity reversal of magnetic fields advected from the accretion flow and resulting in a striped jet \citep{1997ApJ...484..628L,2014MNRAS.440.2185D,2019MNRAS.484.1378G}.
However, such polarity reversal may operate on the scales of $\sim 10^2-10^3 r_{\rm g}/c$, quenching the jets by temporarily flooding the BH surroundings with weakly magnetized plasma, and thus creating long gaps in the jets.
{Although striped jets have been produced in GR force-free simulations \citep{2015MNRAS.446L..61P,2020MNRAS.494.4203M}, evidence for powerful striped jets in GRMHD simulations has been sparse \citep[e.g.,][]{2021MNRAS.508.1241C}.}

The magnetic flux eruptions offer an alternative scenario for introducing a strong perturbation yet preserving the jet core.
The strongest perturbations are expected for the deepest eruptions, which operate over wide azimuthal sectors ($\Delta\phi \sim 1$) and shed $\sim 50\%$ of the initial $\Phi_{\rm BH}$ (magnetic flux drains across the entire BH horizon, diluting magnetic pressure until {the reconnection layer} collapses under the accretion flow; \citealt{2021MNRAS.508.1241C}); which have been obtained in simulations of geometrically thick accretion flows \citep{2011MNRAS.418L..79T,2022ApJ...924L..32R}, in which radiative cooling is not important, and which are thus characteristic for BHs accreting at very low Eddington rates {\citep[e.g.,][]{2025arXiv250508855N}}.
Such deep eruptions can temporarily reduce the jet power (Poynting flux) by factor $\sim 4$.
In geometrically thin accretion flows, in which radiative cooling is important (at high sub-Eddington rates), eruptions are more localized azimuthally ($\Delta\phi \ll 1$) and thus shallow \citep{2022ApJ...935L...1L,2022cosp...44.1769M} -- such eruptions are not expected to result in significant jet distortions, {yet they may be important in modulating jet power and dissipation}.
At super-Eddington rates, accretion flows are geometrically thick, but it is unclear whether magnetic saturation and deep flux eruptions can be realized \citep[e.g.,][]{2026ApJ..1001..138Z}.

\subsection{Magnetic flux tubes and bypasses}

Magnetic flux tubes ejected during magnetic flux eruptions convey relativistically hot plasma from the equatorial reconnection sites \citep{2022ApJ...924L..32R}, most likely also a population of non-thermally accelerated energetic particles that can potentially produce high-energy gamma-ray flares \citep{2023ApJ...943L..29H}.
Additional heating and acceleration can be affected by magnetic Rayleigh-Taylor instability during the interaction between the highly magnetized flux tubes and the weakly magnetized accretion flow \citep{2023PhRvR...5d3023Z}.
As at least some ejected flux tubes are re-advected onto the BH to form magnetic bypasses along the jets.
Such bypasses might convey enough energetic particles (with energies limited by radiative cooling) to produce synchrotron emission at the levels competing with the jet spine (which is expected to be highly sensitive to the viewing angle, since bypasses belong to the mildly relativistic jet sheath).
If resolved across the jet, such emission would appear clearly asymmetric, but not one-sided due to gentle helical twist.
The bypasses are aligned well enough to produce a highly polarized synchrotron signal.
At optically-thin frequencies, for most observers the radiation electric vectors should align perpendicular to the projected jet direction, {a helical distortion may result in rotation of polarization angle \citep{2017Galax...5...64N}}.
However, we found that plasma temperatures along the bypass lines are rather low at $T \sim 10^{-2}$, not elevated in comparison to the jet sheath. Additional processes of plasma heating or particle acceleration would be required for the bypasses to be bright sources of radiation.

\cite{2025ApJ...983...77T} presented dynamical ray-traced synchrotron images of MAD jets from 3D GRMHD simulations for a viewing angle of $17^\circ$ at the frequency of $86\,{\rm GHz}$.
They demonstrated that a BH flux eruption is followed by a clear reduction of the jet image width propagating along the jet at projected distances up to $\sim 100 r_{\rm g}$.
This suggests that magnetic bypasses, expected to provide additional pressure from their poloidal field to confine the jet spine, do not contribute to that emission.

At optically-thick frequencies, weakly magnetized bypasses would form an asymmetric Faraday screen, resulting in rotation measure gradients across the jet \citep{2010ApJ...725..750B}.
Rotation measure gradients across the jet have been detected in many AGN jets, including 3C 273 at the 10-pc scale \citep{2002PASJ...54L..39A}, 3C 120 at pc scale \citep{2008ApJ...681L..69G}, M87 at sub-kpc scale \citep{2021ApJ...923L...5P}.
A characteristic signature of gently helical jet distortion would be an alternating sign of rotation measure gradient along the jet.
However, rotation measure maps based on overall plasma density $\rho$ do not show locally elevated values.
Ejected magnetic flux tubes or bypasses could still form a Faraday screen if loaded by copious electron-positron pairs.
The overall pair content of relativistic jets has been estimated as $n_{\rm e}/n_{\rm p} \sim 20$ \citep{2020MNRAS.499.3749S}.

{Relatively prompt re-advection of ejected magnetic flux is enabled by a limited range ($\sim 32 r_{\rm g}$) of flux tubes ejected against a relatively dense accretion flow from a massive torus.
Longer simulations of more extended, less dense Bondi accretion flows show that the range of such magnetic feedback ({also} termed magneto-convection; \citealt{2022MNRAS.511.2040B}) grows systematically in time \citep{2025ApJ...978..148G,2025ApJ...991...89C}.}

\subsection{Geometry and structure of relativistic jets}

{Our analysis of the geometric shape of the jet (Figure \ref{fig_centroid_spine}), intrinsic distribution of the Poynting flux (Figures \ref{fig_xzmaps2}, \ref{fig_zetaprofs_Sr}), and consistency of the intrinsic jet structure with axisymmetric models (Appendix \ref{sec_res_poloidal_lines}) reveals that the entire jet spine is re-oriented by $\Delta\theta_{\rm jet} \sim 4^\circ$.
{This is consistent with long-term tilt values of $\sim 3^\circ - 5^\circ$ measured at the jet base at $10 r_{\rm g}$ \citep{2025PhRvD.112f3013C}.}
Such re-orientation indicates that magnetic flux eruptions have a more global impact on the jet than generation of waves along the jet sheath \citep{2023ApJ...959L...3D}.
Re-orientation} can strongly modulate the brightness of radiation received by blazar observers, located at small viewing angles $\theta_{\rm obs} \lesssim \arcsin(1/u^t)$.
Even for a modest Lorentz factor of $u^t \simeq 7$ that can be reached by $r \sim 10^3 r_{\rm g}$ (see Figure \ref{fig_zetaprofs_ut}), this means $u^t\,\Delta\theta_{\rm jet} \sim 0.5$.
For $\Delta\theta_{\rm obs} \sim \Delta\theta_{\rm jet} \simeq 0.5/u^t$,
a Doppler factor $\mathcal{D} = [u^t(1-v^r\cos\theta_{\rm obs})]^{-1}$ may change by $\sim 60\%$,
and the apparent luminosity $L \propto \mathcal{D}^4$ by factor $\sim 6$.

Strong modulation of the jet power associated with a helical distortion are expected to evolve over large distances along the jet.
Such evolution cannot be studied with these simulations because of decreasing resolution of a logarithmic radial grid.
Only moderate modulation of the bulk Lorentz factor is found, insufficient to trigger efficient dissipation at distances of $\sim 10^4 r_{\rm g}$ via internal shocks \citep{2001MNRAS.325.1559S}.
The efficiency of relativistic magnetic acceleration --- conversion of $\sigma$ to $u^t$ (with $\sigma < 1$ important for dissipative shocks; \citealt{2015MNRAS.450..183S}) --- is low by the distance of $10^3 r_{\rm g}$ \citep{2024MNRAS.533..254S}.

Basic predictions of semi-analytical relativistic MHD models \citep[e.g.,][]{2009ApJ...698.1570L} can be reproduced {in analysis of the intrinsic structure of the jet spine, measured from the physical jet axis, which is geometrically shifted from the coordinate axis (Section \ref{sec_res_struct}, Appendix \ref{sec_res_poloidal_lines}).}

We find low-amplitude structures along the jet core (in magnetic field $B^z$ -- see the green dashed boxes in the middle panel of Figure \ref{fig_xzmaps_Bz}; also in the Lorentz factor $u^t$ -- see the left panel of Figure \ref{fig_xzmaps2} at $-20 r_{\rm g} < x < 40 r_{\rm g}$ and $z > 580 r_{\rm g}$).
They appear to be quasi-periodic, with a period of $\lambda_z \simeq 45 r_{\rm g}$ estimated for $580 < z/r_{\rm g} < 1000$, slightly higher than the magnetic pitch $\mathcal{P} \simeq 36 r_{\rm g}$ (see Section \ref{sec_res_struct}, Figure \ref{fig_zetaprofs_pitch})\footnote{We note that the magnetic pitch $\mathcal{P}$ is well resolved -- at the distance of $r \sim 10^3 r_{\rm g}$ the linear resolution of this simulation is ${\rm d}r \simeq r\,{\rm d}\theta \sim 1.36 r_{\rm g}$, hence $\mathcal{P} \sim 26 {\rm d}r$.}.
We tentatively interpret them as the current-driven instability (CDI) kink modes \citep{2000A&A...355..818A}, {noting that in local simulations of cylindrical columns such modes expand outwards from the poloidal core \citep{2019ApJ...884...39B,2022ApJ...931..137O}.}
Using the prescription of \cite{2017MNRAS.468.4635S} (see also \citealt{1999MNRAS.308.1006L}), we estimate the CDI growth time scale\footnote{According to \cite{2017MNRAS.468.4635S}, the CDI growth frequency is $\omega_{\rm CDI} \sim 0.17\alpha/(\Gamma R_0)$ (see their Eq. 25) for the poloidal magnetic field profile $B^z(R) = B_0[1+(R/R_0)^2]^{-\alpha}$ (see their Eq. 7). At the distance of $r_0 \simeq 830 r_{\rm g}$ we estimate the intrinsic poloidal field profile as consistent with $\alpha \simeq 0.4-0.5$ and $R_0 \simeq 26 r_{\rm g}\,(r/r_0)^{1/2}$, with the bulk Lorentz factor $\Gamma \simeq 2.7$. This yields $\omega_{\rm CDI} \equiv \tau_{\rm CDI}^{-1} \sim 0.028/R_{\rm 0}$, hence $\tau_{\rm CDI} \sim (26/0.028)(r/830 r_{\rm g})^{1/2}\,r_{\rm g}$.} as $\tau_{\rm CDI} \simeq (r\times 10^3 r_{\rm g})^{1/2}$.
This becomes comparable with the adiabatic time scale\footnote{The jet spine expands paraboloidally with $R \propto r^{1/2}$ (see the upper panel of Figure \ref{fig_poloidal_lines}). The toroidal magnetic field driving the CDI decreases like $B^\phi \propto R^{-1} \propto r^{-1/2}$. With ${\rm d}r/{\rm d}t = v^r$, one has $\tau_{\rm ad} \equiv B^\phi/|{\rm d}B^\phi/{\rm d}t| \simeq 2r/v^r$ \citep{2008A&A...492..621M}.} $\tau_{\rm ad} \sim 2r/v^r$ at the distance of $r_{\rm CDI} \sim 250 r_{\rm g}$ where $v^r \simeq 0.95$, hence the CDI modes should have a chance to grow for $r > r_{\rm CDI}$ where $\tau_{\rm CDI} < \tau_{\rm ad}$.

We find that $\sim (90-95)\%$ of the jet Poynting flux $S^r$ is contained within the jet spine, defined instantaneously by the cold magnetization $\sigma_{\rm c} > 1$.
This is different from the results reported by \cite{2025ApJ...979..199S}, based on GRMHD simulations with uniform resistivity in 2D axisymmetry, that even $\sim 50\%$ of total Poynting flux outflowing from the vicinity of magnetized accreting black hole could be carried by the jet sheath.
Formally, their definition of spine/sheath boundary is different -- based on time-averaged maps.
Qualitatively, their sheath region appears more dynamic with plenty substructures.
Whether this could be due to either resistivity or axisymmetry requires additional investigation.

\subsection{Generalizing the results of this work}

{This work is limited to analysis of a single case of magnetic flux eruption and re-accumulation simulated at exceptionally high resolution.
The resolution study of \cite{2024MNRAS.533..254S} showed that general features of magnetically saturated (MAD) geometrically thick accretion flows can be obtained at lower resolutions.
Previous simulations performed at lower resolutions extended to much longer times and produced many eruptions of varying depths \citep[e.g.,][]{2022MNRAS.511.3795N}.
Previous studies explored the effect of BH spin $a$ on MADs \citep[e.g.,][]{2022MNRAS.511.3795N,2025ApJ...983...77T}, and a new high-resolution survey of $a$ is ongoing.
The effect of initial torus tilt has been investigated by \cite{2025PhRvD.112f3013C} --- they found similar wobbles at the jet base.
The effect of shallow eruptions from magnetically truncated geometrically thin accretion disks \citep{2022ApJ...935L...1L,2022cosp...44.1769M} can also be investigated.}

\section{Conclusions}
\label{sec_conc}

We have analyzed selected results from the high-resolution 3D ideal GRMHD numerical simulation (produced with the H-AMR code) of magnetically saturated (MAD state) geometrically thick accretion flow onto a Kerr BH with high spin ($a = 15/16$), launching powerful relativistic jets followed to the distance of $2000 r_{\rm g}$ \citep{2022ApJ...924L..32R}.
We analyzed the geometry of relativistically magnetized ($\sigma > 1$) jet spine distorted after a reconnection-driven eruption of BH magnetic flux.
In addition to waves modulating the jet sheath \citep{2023ApJ...959L...3D}, a global helical distortion develops along the jet section of reduced power (mainly the Poynting flux), its length ($\Delta z \sim 100 r_{\rm g}$) clearly shorter than the eruption time scale ($\Delta t \sim 200 r_{\rm g}/c$).
It is followed by a re-powered section of jet spine tilted by $\Delta\theta \sim 4^\circ$, implying a significant change of Doppler factor for observers of certain blazars {(cf. \citealt{2025PhRvD.112f3013C})}.
The jet spine remains contiguous due to a fixed amount of (largely poloidal) magnetic flux $\Phi_{\rm BH,sp}$ passing from the BH horizon along the jet spine to the outer boundary.
Other components of magnetic flux are modulated by the cycle of BH flux eruption and accumulation.
The magnetic flux ejected during the eruption consists of: (1) tubes orbiting as hotspots in the equatorial plane, passing through the jet sheath ($\sigma < 1$), and joining the jet spine along the upstream edge of the helical distortion; and (2) bypasses re-advected onto the BH and re-powered spine, passing closely along the helical distortion but still within the jet sheath (filling the space around the distorted spine, effectively rounding it up towards a symmetric cone), indicating that they were relatively quickly re-loaded with cold plasma to $\sigma < 1$ and $T \sim 10^{-2}$.
The downstream edge of the helical distortion appears to obliquely impact the jet sheath, with signs of plasma heating on the spine side.
The intrinsic structure of the jet spine is consistent with axisymmetric semi-analytic models of the acceleration-collimation zone before conversion of magnetic-to-kinetic energy and lateral differentiation.
The intrinsic jet shows weak quasi-periodic structures, with spacing comparable to the uniform magnetic pitch, that could be signatures of the current-driven kink modes.
Such strong jet distortions can be produced by deep magnetic flux eruptions from black holes fed by geometrically thick accretion flows, generally expected either at very low or at super Eddington rates.
They could explain asymmetric superluminal knots or jet re-orientation in low-luminosity AGNs like M87 and associated with them less luminous blazars of the BL Lac type.

\paragraph{Acknowledgments.}
We thank Gibwa Musoke, Sebastiano
von Fellenberg {and Koushik Chatterjee} for helpful comments.
We acknowledge Poland's high-performance Infrastructure PLGrid (HPC Centers: ACK Cyfronet AGH, PCSS, CI TASK, WCSS) for providing computer facilities and support within computational grant no. PLG/2024/017013 and PLG/2025/018500;
and the Nicolaus Copernicus Astronomical Center for providing the Chuck cluster.
This work was supported in part by the Polish National Science Centre grants 2021/41/B/ST9/04306 and 2024/53/B/ST9/03747.
This research was supported in part by grant NSF PHY-2309135 to the Kavli Institute for Theoretical Physics (KITP).

AP is supported by NASA grant 80NSSC22K1054,  and facilitated by Multimessenger Plasma Physics Center (MPPC), NSF grant No. PHY-2206607.
AP additionally acknowledges support by an Alfred P. Sloan Fellowship, and a Packard Foundation Fellowship in Science and Engineering.
BR is supported by the Natural Sciences \& Engineering Research Council of Canada (NSERC; funding reference number 568580), and the Canadian Space Agency (23JWGO2A01).
AP and BR are supported by the Simons Foundation (grant 00001470), and KN acknowledges their hospitality.

Part of the simulations were performed on the DOE OLCF Summit under the allocation AST198.
This research used resources of the Oak Ridge Leadership Computing Facility at the Oak Ridge National Laboratory, which is supported by the Office of Science of the U.S. Department of Energy under Contract No. DE-AC05-00OR22725.
The computational resources and services used in this work were also partially provided by facilities supported by the VSC (Flemish Supercomputer Center), funded by the Research Foundation Flanders (FWO) and the Flemish Government Department EWI, by Compute Ontario and the Digital Research Alliance of Canada (alliancecan.ca) compute allocation {\tt rrg-ripperda}, and the CCA at the Flatiron Institute supported by the Simons Foundation.


~\clearpage
\begin{appendix}

\section{Intrinsic poloidal field lines and magnetic pitch}
\label{sec_res_poloidal_lines}

\begin{figure}[h]
\includegraphics[width=\columnwidth]{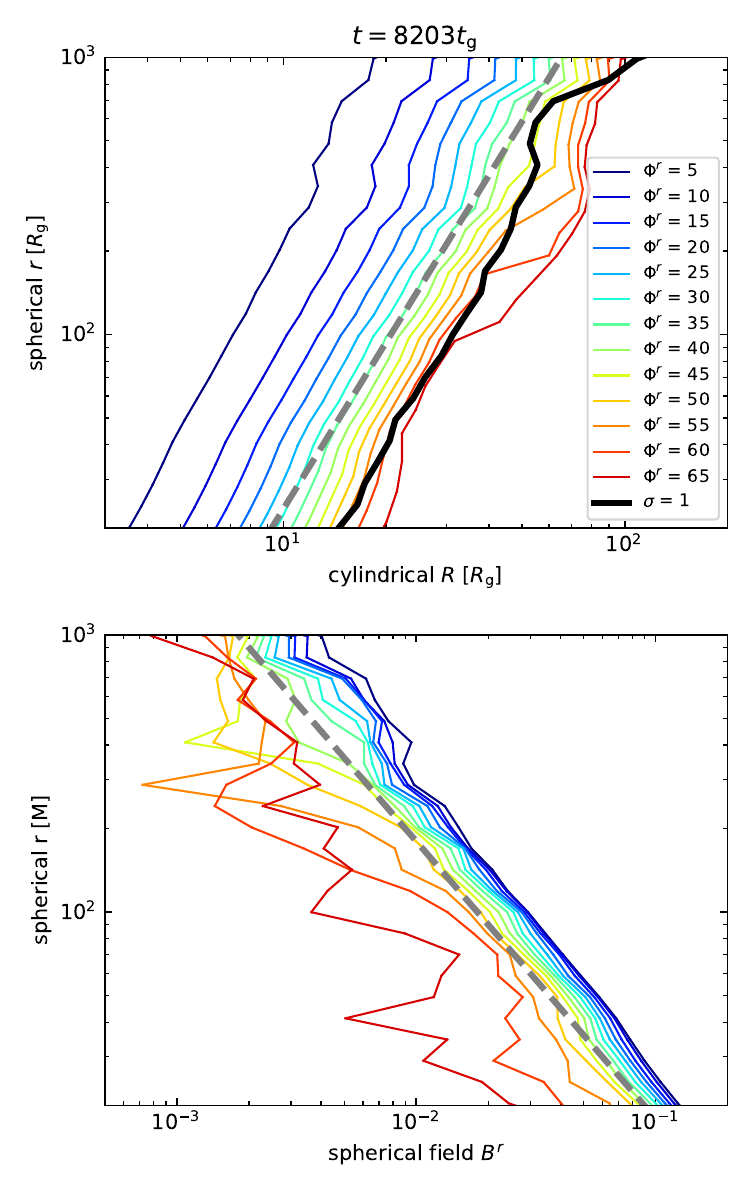}
\caption{Upper panel: intrinsic poloidal field lines, contours of spherical-radial magnetic flux $\Phi^r$, functions of the intrinsic cylindrical radius $R$; the thick black solid line indicates the jet spine boundary (contour of $\sigma = 1$); the thick gray dashed line marks a parabola $R\propto r^{1/2}$.
Lower panel: strength of spherical-radial magnetic field $B^r$ along the same intrinsic poloidal field lines; the thick gray dashed line marks a power law $B^r \propto r^{-1}$.}
\label{fig_poloidal_lines}
\end{figure}

Figure \ref{fig_poloidal_lines} shows intrinsic poloidal field lines, i.e., contours of the spherical-radial magnetic flux integrated from the jet axis
\be
\Phi^r(\zeta) = \pi r^2 \int_0^\zeta {\rm d}\zeta'\, \sin(2\zeta')\, B^r(\zeta')\,,
\ee
as functions of cylindrical radius $R = r\sin\theta$.
Within the jet spine, the poloidal lines are consistent with parabolas $R \propto r^{1/2}$ {(dashed line)}.
The strength of spherical-radial magnetic field $B^r$ along those lines is consistent with $B^r \propto r^{-1}$ {(dashed line)} within the jet spine.

\begin{figure}[h]
\includegraphics[width=\columnwidth]{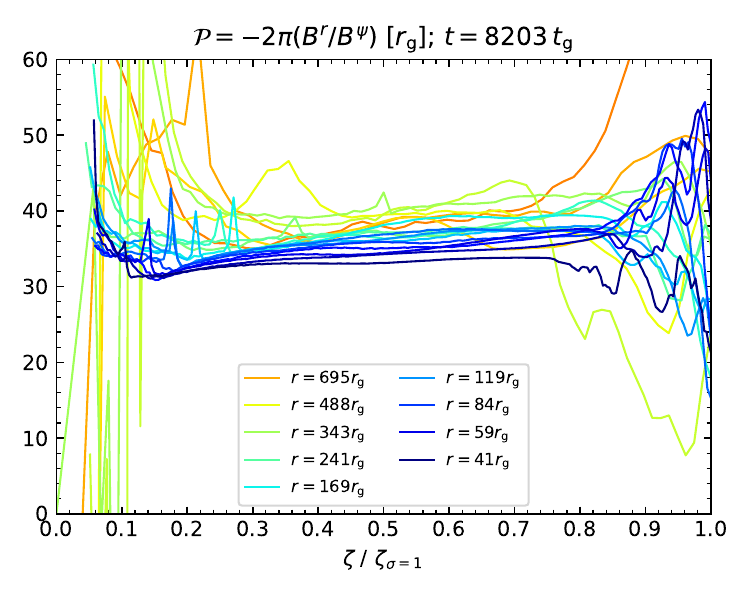}
\caption{Intrinsic profiles of the magnetic pitch $\mathcal{P}$ in units of $r_{\rm g}$,
functions of the polar angle $\zeta$ normalized to the spine boundary $\zeta_{\sigma=1}$, compared for many values of the spherical radius $r$.}
\label{fig_zetaprofs_pitch}
\end{figure}

Figure \ref{fig_zetaprofs_pitch} presents the intrinsic magnetic pitch $\mathcal{P} = -2\pi(B^r/B^\psi)$ (in units of length) as functions of $\zeta$ for multiple values of $r$.
Defined in this way, it represents the pitch length of individual magnetic field lines (i.e., radial distance $\Delta r$ covered over a full toroidal rotation $\Delta\psi = 2\pi$).
It is roughly uniform in both $\zeta$ and $r$ with $\mathcal{P} \sim 36 r_{\rm g}$; trends of $\mathcal{P}$ increasing slightly with increasing $\zeta$ and increasing $r$ can be noticed.
Across the jet spine, magnetic field lines are thus characterized by roughly uniform pitch.

\section{Oblique spine-sheath interface}
\label{sec_res_impact}

\begin{figure*}
\includegraphics[width=\textwidth]{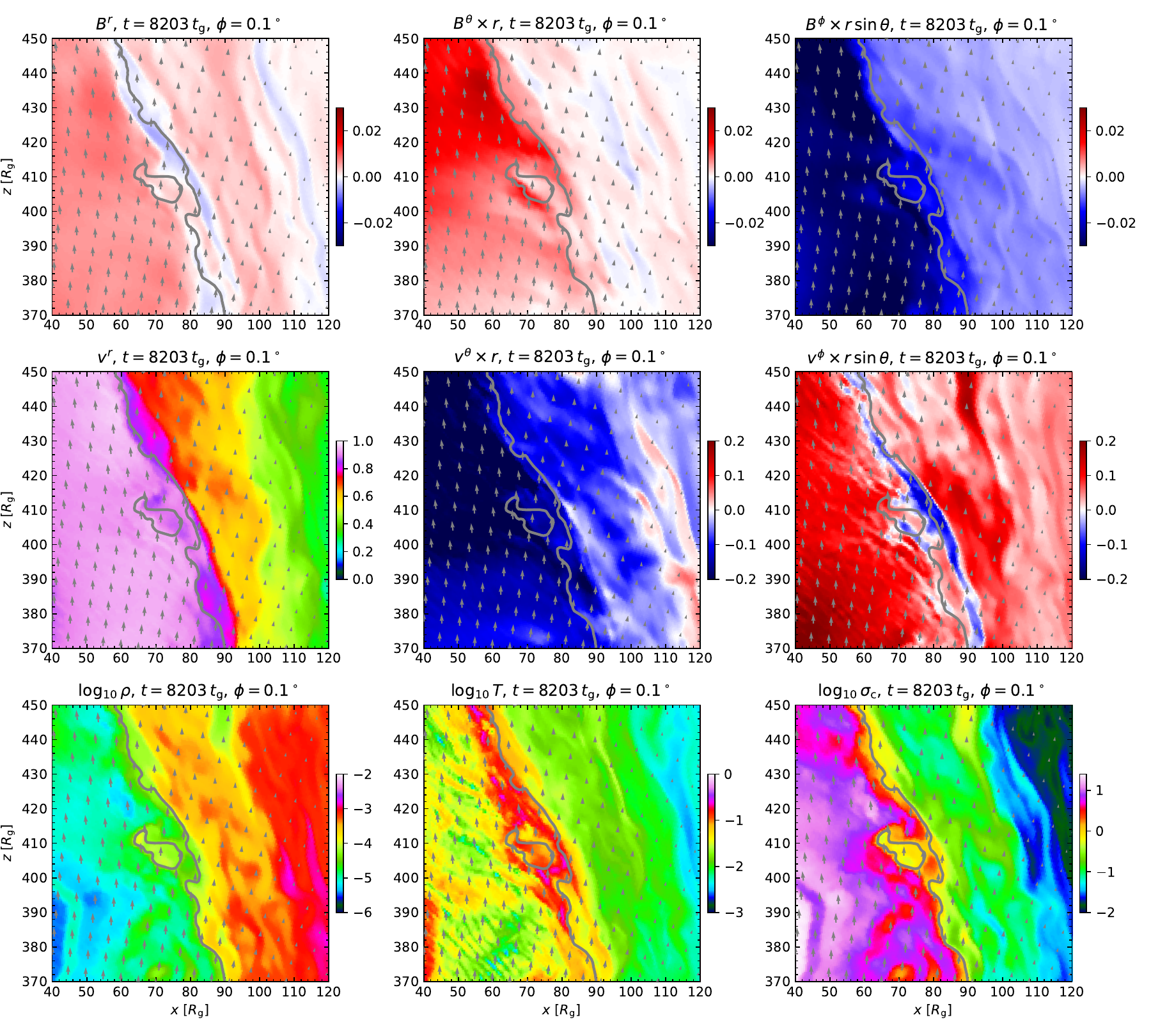}
\caption{Maps in the $(x,z)$ coordinates of the region (indicated also in the middle panel of Figure \ref{fig_xzmaps_Bz} and in Figure \ref{fig_xzmaps2}) where the reoriented highly magnetized ($\sigma_{\rm c} > 1$) jet spine (lower left side) impacts the weakly magnetized ($\sigma_{\rm c} < 1$) jet sheath (upper right side), forming an oblique interface.
Each panel shows a color map for different parameter indicated in the panel title.
The gray arrows indicate the in-plane velocity field $(v^x,v^z)$.
The gray solid line marks the contour of $\sigma_{\rm c} = 1$.}
\label{fig_xzmaps_shock}
\end{figure*}

Figure \ref{fig_xzmaps_shock} presents maps of multiple parameters zoomed on the region (indicated with a gray box in the middle panel of Figure \ref{fig_xzmaps_Bz} and in Figure \ref{fig_xzmaps2}) where an oblique front of the re-oriented and re-powered jet spine ($\sigma > 1$) interacts with the sheath medium ($\sigma < 1$).
A sharp contrast between spine and sheath can be seen in parameters other than magnetization: the jet spine is characterized by relativistic radial velocity $v^r > 0.8$ and low plasma density $\log_{10}\rho < -4$; the sheath is characterized by mildly relativistic radial velocity $v^r < 0.7$ and high plasma density.
The velocity field is dominated by the radial component $v^r$ both in the spine and sheath, suggesting a direct impact of the faster spine on the slower sheath.
Plasma temperature is clearly elevated ($\log_{10}T > -1$) over a broad patch ($\Delta r \sim 20 r_{\rm g}$) on the spine side.
In addition, the radial magnetic field $B^r$ and the azimuthal velocity $v^\phi$ are reversed within a narrow band crossing the spine-sheath boundary at a small angle.

The oblique interface between the jet spine and sheath is most likely a contact discontinuity, for which no dissipation is expected.
A curious feature is the elevated temperature $T > 0.1$ over a $\Delta r \sim 15 r_{\rm g}$ band on the spine side.
The origin of this excess heat (whether physical or numerical) is not clear.

\end{appendix}


\begin{thebibliography}{}

\bibitem[Ackermann et~al.(2016)]{2016ApJ...824L..20A}
Ackermann, M., Anantua, R., Asano, K., et~al.\ 2016, \apjl, 824, 2, L20

\bibitem[Algaba et~al.(2024)]{2024A&A...692A.140A}
{Algaba, J.~C., Balokovi{\'c}, M., Chandra, S., et~al.\ 2024, \aap, 692, A140}

\bibitem[Appl et~al.(2000)]{2000A&A...355..818A}
{Appl, S., Lery, T., \& Baty, H.\ 2000, \aap, 355, 818}

\bibitem[Asada et~al.(2002)]{2002PASJ...54L..39A}
Asada, K., Inoue, M., Uchida, Y., et~al.\ 2002, \pasj, 54, L39

\bibitem[Begelman et~al.(2022)]{2022MNRAS.511.2040B}
{Begelman, M.~C., Scepi, N., \& Dexter, J.\ 2022, \mnras, 511, 2, 2040}

\bibitem[Beskin \& Nokhrina(2009)]{2009MNRAS.397.1486B}
Beskin, V.~S. \& Nokhrina, E.~E.\ 2009, \mnras, 397, 3, 1486

\bibitem[Blandford(1976)]{1976MNRAS.176..465B}
Blandford, R.~D.\ 1976, \mnras, 176, 465

\bibitem[Blandford \& Znajek(1977)]{1977MNRAS.179..433B}
Blandford, R.~D. \& Znajek, R.~L.\ 1977, \mnras, 179, 433

\bibitem[Blandford et~al.(2019)]{2019ARA&A..57..467B}
Blandford, R., Meier, D., \& Readhead, A.\ 2019, \araa, 57, 467

\bibitem[Bridle et~al.(1994)]{1994AJ....108..766B} Bridle, A.~H., Hough, D.~H., Lonsdale, C.~J., et~al.\ 1994, \aj, 108, 766

\bibitem[Broderick \& McKinney(2010)]{2010ApJ...725..750B}
Broderick, A.~E. \& McKinney, J.~C.\ 2010, \apj, 725, 1, 750

\bibitem[Bromberg et~al.(2019)]{2019ApJ...884...39B}
Bromberg, O., Singh, C.~B., Davelaar, J., et~al.\ 2019, \apj, 884, 1, 39

\bibitem[Camenzind(1989)]{1989ASSL..156..129C}
Camenzind, M.\ 1989, Accretion Disks and Magnetic Fields in Astrophysics, 156, 129

\bibitem[Chashkina et~al.(2021)]{2021MNRAS.508.1241C}
Chashkina, A., Bromberg, O., \& Levinson, A.\ 2021, \mnras, 508, 1241

\bibitem[Chatterjee et~al.(2025)]{2025PhRvD.112f3013C}
{Chatterjee, K., Kaaz, N., Liska, M., et~al.\ 2025, \prd, 112, 6, 063013}

\bibitem[Cho \& Narayan(2025)]{2025ApJ...991...89C}
{Cho, H. \& Narayan, R.\ 2025, \apj, 991, 1, 89}

\bibitem[Davelaar et~al.(2018)]{2018A&A...612A..34D}
Davelaar, J., Mo{\'s}cibrodzka, M., Bronzwaer, T., et~al.\ 2018, \aap, 612, A34

\bibitem[Davelaar et~al.(2023)]{2023ApJ...959L...3D}
Davelaar, J., Ripperda, B., Sironi, L., et~al.\ 2023, \apjl, 959, 1, L3

\bibitem[Dexter et~al.(2014)]{2014MNRAS.440.2185D} Dexter, J., McKinney, J.~C., Markoff, S., et~al.\ 2014, \mnras, 440, 3, 2185

\bibitem[Fuentes et~al.(2023)]{2023NatAs...7.1359F} Fuentes, A., G{\'o}mez, J.~L., Mart{\'\i}, J.~M., et~al.\ 2023, Nature Astronomy, 7, 1359

\bibitem[Galishnikova et~al.(2025)]{2025ApJ...978..148G}
{Galishnikova, A., Philippov, A., Quataert, E., et~al.\ 2025, \apj, 978, 2, 148}

\bibitem[Giannios et~al.(2009)]{2009MNRAS.395L..29G} Giannios, D., Uzdensky, D.~A., \& Begelman, M.~C.\ 2009, \mnras, 395, 1, L29

\bibitem[Giannios \& Uzdensky(2019)]{2019MNRAS.484.1378G}
Giannios, D. \& Uzdensky, D.~A.\ 2019, \mnras, 484, 1, 1378

\bibitem[G{\'o}mez et~al.(2008)]{2008ApJ...681L..69G}
G{\'o}mez, J.~L., Marscher, A.~P., Jorstad, S.~G., et al.\ 2008, \apjl, 681, 2, L69

\bibitem[Goyal et~al.(2022)]{2022ApJ...927..214G}
Goyal, A., Soida, M., Stawarz, {\L}., et~al.\ 2022, \apj, 927, 2, 214

\bibitem[Hakobyan et~al.(2023)]{2023ApJ...943L..29H}
Hakobyan, H., Ripperda, B., \& Philippov, A.~A.\ 2023, \apjl, 943, 2, L29

\bibitem[Harris et~al.(2009)]{2009ApJ...699..305H} Harris, D.~E., Cheung, C.~C., Stawarz, {\L}., et~al.\ 2009, \apj, 699, 1, 305

\bibitem[Jacquemin-Ide et~al.(2025)]{2025arXiv251025842J}
Jacquemin-Ide, J., Begelman, M.~C., Lowell, B., et~al.\ 2025, arXiv:2510.25842

\bibitem[Jia et~al.(2023)]{2023MNRAS.526.2924J}
Jia, H., Ripperda, B., Quataert, E., et~al.\ 2023, \mnras, 526, 2, 2924

\bibitem[Jorstad et~al.(2001)]{2001ApJS..134..181J} Jorstad, S.~G., Marscher, A.~P., Mattox, J.~R., et~al.\ 2001, \apjs, 134, 2, 181

\bibitem[Jorstad et~al.(2017)]{2017ApJ...846...98J} Jorstad, S.~G., Marscher, A.~P., Morozova, D.~A., et~al.\ 2017, \apj, 846, 2, 98

\bibitem[Kim et~al.(2020)]{2020A&A...640A..69K}
Kim, J.-Y., Krichbaum, T.~P., Broderick, A.~E., et~al.\ 2020, \aap, 640, A69

\bibitem[Liska et~al.(2022)]{2022ApJ...935L...1L}
Liska, M.~T.~P., Musoke, G., Tchekhovskoy, A., et~al.\ 2022, \apjl, 935, 1, L1

\bibitem[Liska et~al.(2022)]{2022ApJS..263...26L}
Liska, M.~T.~P., Chatterjee, K., Issa, D., et~al.\ 2022, \apjs, 263, 2, 26

\bibitem[Lister et~al.(2009)]{2009AJ....137.3718L} Lister, M.~L., Aller, H.~D., Aller, M.~F., et~al.\ 2009, \aj, 137, 3, 3718

\bibitem[Lister et~al.(2016)]{2016AJ....152...12L}
Lister, M.~L., Aller, M.~F., Aller, H.~D., et~al.\ 2016, \aj, 152, 1, 12

\bibitem[Lovelace et~al.(1997)]{1997ApJ...484..628L}
Lovelace, R.~V.~E., Newman, W.~I., \& Romanova, M.~M.\ 1997, \apj, 484, 2, 628

\bibitem[Lyubarskii(1999)]{1999MNRAS.308.1006L}
{Lyubarskii, Y.~E.\ 1999, \mnras, 308, 4, 1006}

\bibitem[Lyubarsky(2009)]{2009ApJ...698.1570L}
Lyubarsky, Y.\ 2009, \apj, 698, 2, 1570

\bibitem[Madejski \& Sikora(2016)]{2016ARA&A..54..725M}
Madejski, G. \& Sikora, M.\ 2016, \araa, 54, 725

\bibitem[Mahlmann et~al.(2020)]{2020MNRAS.494.4203M}
{Mahlmann, J.~F., Levinson, A., \& Aloy, M.~A.\ 2020, \mnras, 494, 3, 4203}

\bibitem[Marscher et~al.(2002)]{2002Natur.417..625M}
Marscher, A.~P., Jorstad, S.~G., G{\'o}mez, J.-L., et al.\ 2002, \nat, 417, 6889, 625

\bibitem[McKinney \& Uzdensky(2012)]{2012MNRAS.419..573M}
McKinney, J.~C. \& Uzdensky, D.~A.\ 2012, \mnras, 419, 1, 573

\bibitem[Mirabel et~al.(1998)]{1998A&A...330L...9M}
Mirabel, I.~F., Dhawan, V., Chaty, S., et~al.\ 1998, \aap, 330, L9

\bibitem[Mirabel \& Rodr{\'\i}guez(1999)]{1999ARA&A..37..409M}
Mirabel, I.~F. \& Rodr{\'\i}guez, L.~F.\ 1999, \araa, 37, 409

\bibitem[Moll et~al.(2008)]{2008A&A...492..621M}
{Moll, R., Spruit, H.~C., \& Obergaulinger, M.\ 2008, \aap, 492, 3, 621}

\bibitem[Musoke et~al.(2022)]{2022cosp...44.1769M}
{Musoke, G., Porth, O., \& Liska, M.\ 2022, 44th COSPAR Scientific Assembly, 44, 1769}

\bibitem[Naethe Motta et~al.(2025)]{2025arXiv250508855N}
{Naethe Motta, P., Jacquemin-Ide, J., Nemmen, R., et~al.\ 2025, arXiv:2505.08855}

\bibitem[Nalewajko \& Sikora(2009)]{2009MNRAS.392.1205N}
Nalewajko, K. \& Sikora, M.\ 2009, \mnras, 392, 1205

\bibitem[Nalewajko et~al.(2011)]{2011MNRAS.413..333N}
Nalewajko, K., Giannios, D., Begelman, M.~C., et~al.\ 2011, \mnras, 413, 1, 333

\bibitem[Nalewajko et~al.(2014)]{2014ApJ...789..161N}
Nalewajko, K., Begelman, M.~C., \& Sikora, M.\ 2014, \apj, 789, 2, 161

\bibitem[Nalewajko(2017)]{2017Galax...5...64N}
{Nalewajko, K.\ 2017, Galaxies, 5, 4, 64}

\bibitem[Nalewajko et~al.(2024)]{2024A&A...692A..37N}
Nalewajko, K., Kapusta, M., \& Janiuk, A.\ 2024, \aap, 692, A37

\bibitem[Narayan et~al.(2003)]{2003PASJ...55L..69N}
Narayan, R., Igumenshchev, I.~V., \& Abramowicz, M.~A.\ 2003, \pasj, 55, L69

\bibitem[Narayan et~al.(2022)]{2022MNRAS.511.3795N}
{Narayan, R., Chael, A., Chatterjee, K., et~al.\ 2022, \mnras, 511, 3, 3795}

\bibitem[O'Connor et~al.(2023)]{2023SciA....9I1405O}
O'Connor, B., Troja, E., Ryan, G., et~al.\ 2023, Science Advances, 9, 23, eadi1405

\bibitem[Ortu{\~n}o-Mac{\'\i}as et~al.(2022)]{2022ApJ...931..137O}
{Ortu{\~n}o-Mac{\'\i}as, J., Nalewajko, K., Uzdensky, D.~A., et~al.\ 2022, \apj, 931, 2, 137}

\bibitem[Parfrey et~al.(2015)]{2015MNRAS.446L..61P}
{Parfrey, K., Giannios, D., \& Beloborodov, A.~M.\ 2015, \mnras, 446, L61}

\bibitem[Pasetto et~al.(2021)]{2021ApJ...923L...5P} Pasetto, A., Carrasco-Gonz{\'a}lez, C., G{\'o}mez, J.~L., et al.\ 2021, \apjl, 923, 1, L5

\bibitem[Piran(2004)]{2004RvMP...76.1143P}
Piran, T.\ 2004, RvMP, 76, 1143

\bibitem[Rees \& Meszaros(1994)]{1994ApJ...430L..93R}
Rees, M.~J. \& Meszaros, P.\ 1994, \apjl, 430, L93

\bibitem[Ripperda et~al.(2020)]{2020ApJ...900..100R}
Ripperda, B., Bacchini, F., \& Philippov, A.~A.\ 2020, \apj, 900, 2, 100

\bibitem[Ripperda et~al.(2022)]{2022ApJ...924L..32R}
Ripperda, B., Liska, M., Chatterjee, K., et~al.\ 2022, \apjl, 924, L32

\bibitem[Ro et~al.(2026)]{2026ApJ...999..169R}
Ro, H., Kino, M., Hada, K., et~al.\ 2026, \apj, 999, 2, 169

\bibitem[Rybicki \& Lightman(1986)]{1986rpa..book.....R}
Rybicki, G.~B. \& Lightman, A.~P.\ 1986, Radiative Processes in Astrophysics, pp. 400. ISBN 0-471-82759-2. Wiley-VCH

\bibitem[Salas et~al.(2024)]{2024MNRAS.533..254S}
Salas, L.~D.~S., Musoke, G., Chatterjee, K., et~al.\ 2024, \mnras, 533, 254

\bibitem[Sikora et~al.(2020)]{2020MNRAS.499.3749S}
Sikora, M., Nalewajko, K., \& Madejski, G.~M.\ 2020, \mnras, 499, 3, 3749

\bibitem[Sironi et~al.(2015)]{2015MNRAS.450..183S}
Sironi, L., Petropoulou, M., \& Giannios, D.\ 2015, \mnras, 450, 183

\bibitem[Sobacchi et~al.(2017)]{2017MNRAS.468.4635S}
{Sobacchi, E., Lyubarsky, Y.~E., \& Sormani, M.~C.\ 2017, \mnras, 468, 4, 4635}

\bibitem[Spada et~al.(2001)]{2001MNRAS.325.1559S}
Spada, M., Ghisellini, G., Lazzati, D., et~al.\ 2001, \mnras, 325, 1559

\bibitem[Sridhar et~al.(2025)]{2025ApJ...979..199S}
Sridhar, N., Ripperda, B., Sironi, L., et~al.\ 2025, \apj, 979, 2, 199

\bibitem[Tchekhovskoy et~al.(2011)]{2011MNRAS.418L..79T}
Tchekhovskoy, A., Narayan, R., \& McKinney, J.~C.\ 2011, \mnras, 418, L79

\bibitem[Tsunetoe et~al.(2025)]{2025ApJ...983...77T}
Tsunetoe, Y., Narayan, R., \& Ricarte, A.\ 2025, \apj, 983, 1, 77

\bibitem[Weaver et~al.(2022)]{2022ApJS..260...12W} Weaver, Z.~R., Jorstad, S.~G., Marscher, A.~P., et~al.\ 2022, \apjs, 260, 1, 12

\bibitem[Zhang \& Yan(2011)]{2011ApJ...726...90Z}
Zhang, B. \& Yan, H.\ 2011, \apj, 726, 2, 90

\bibitem[Zhang et al.(2026)]{2026ApJ..1001..138Z}
{Zhang, L., Stone, J.~M., White, C.~J., et~al.\ 2026, \apj, 1001, 2, 138}

\bibitem[Zhdankin et~al.(2023)]{2023PhRvR...5d3023Z}
Zhdankin, V., Ripperda, B., \& Philippov, A.~A.\ 2023, Physical Review Research, 5, 4, 043023

\end{thebibliography}
\end{document}